\documentclass[%
reprint,
 amsmath,amssymb,
 aps,
 prl,
]{revtex4-2}

\usepackage{graphicx}
\usepackage{dcolumn}
\usepackage{bm}

\usepackage{siunitx}
\usepackage[version=4]{mhchem}
\usepackage{physics}
\usepackage{bm}
\usepackage{CJK}

\begin{document}

\preprint{APS/123-QED}

\title{Instability and Momentum Bifurcation of a molecular BEC \\ in a Shaken Lattice with Exotic Dispersion}

\begin{CJK*}{UTF8}{gbsn} 

\author{Kaiyue Wang (王凯越), Feng Xiong (熊风), Yun Long (龙云), Yun Ma (马芸), Colin V. Parker}
\affiliation{
School of Physics, Georgia Institute of Technology, Atlanta, Georgia 30332, USA
}%

\date{\today}

\begin{abstract}
We place a molecular Bose-Einstein condensate in a 1D shaken lattice with a Floquet-engineered dispersion, and observe the dynamics in both position and momentum space. At the initial condition of zero momentum, our engineered dispersion is inverted, and therefore unstable. We observe that the condensate is destabilized by the lattice shaking as expected, but rather than decaying incoherently or producing jets, as in other unstable condensates, under our conditions the condensate bifurcates into two portions in momentum space, with each portion subsequently following semi-classical trajectories that suffer minimal spreading in momentum space as they evolve. We can model the evolution with a Gross-Pitaevskii equation, which suggests the initial bifurcation is facilitate by a nearly linear ``inverted V''-shaped dispersion at the zone center, while the lack of spreading in momentum space is facilitated by interactions, as in a soliton. We propose that this relatively clean bifurcation in momentum space has applications for counter-diabatic preparation of exotic ground states in many-body quantum simulation schemes. 
\end{abstract}

\maketitle
\end{CJK*} 

Degenerate quantum gases of ultracold atoms have emerged as powerful simulators of both equilibrium and non-equilibrium properties. One method of non-equilibrium simulation is to prepare Bose-Einstein condensates (BECs) with initial conditions far from the ground state and study the resulting dynamics. In some cases, the nominally unstable point can in fact be at least quasi-stable, such as the recently observed soliton in an inverted band \cite{mitchell2021floquet} or many-body scar states \cite{serbyn2021quantum} which lead to anomalously long lifetimes for spin helices\cite{jepsen2022long}. In other cases, dramatic types of decay can be observed such as the so-called Bose-Nova \cite{lahaye2008d} and Bose Fireworks \cite{clark2017collective}. However, preparation of initial conditions with macroscopic occupation of multiple points in phase space can be challenging using adiabatic preparation\cite{clark2016universal,yao2022domain}, if the system is not stable over the required ramp time. An alternative is to use counter-diabatic methods to move dynamically across the transition\cite{sels_minimizing_2017}. In this work we show how a shaken lattice can be used to rapidly prepare ``bifurcated'' condensates with macroscopic occupation of two points in phase space.

The shaken lattice is a well-known technique capable of modifying the energy-momentum dispersion relation of the system's effective Hamiltonian \cite{eckardt2017colloquium,weitenberg2021tailoring} and has been used to study dynamics by generating artificial interactions \cite{zahn2022formation,struck2014spin,struck2012tunable}, gauge fields \cite{jotzu2014experimental, yao2022domain}, or band topologies \cite{sandholzer2022floquet}. By coupling the lowest two bands with near-resonant periodic driving, one of the hybrid bands features two stable minima at tunable quasimomentum alongside the unstable Brillouin zone center, while stronger off-resonant shaking yields an inverted band and negative mass. These exotic band shape with tunable balance and separation in minima can simulate phase transitions and domain dynamics \cite{parker2013direct,clark2016universal,anderson2017direct,yao2022domain,song2022realizing}. This feature is proposed to be used for generating complex Fermi surfaces and unconventional fermionic pairing \cite{kelecs2017effective, zhang2015chiral,kawamura2022proposed}, particularly the Fulde--Ferrell--Larkin--Ovchinnikov (FFLO) phases \cite{zheng2015floquet,zheng2016fulde}, which are of great interest but challenging to observe \cite{kinnunen2018fulde,liao2010spin,partridge2006pairing,olsen2015phase}. Many of the proposed schemes involve non-equilibrium loading in a shaken optical lattice to simulate spin imbalance, as we demonstrate here.

\begin{figure}[h!]
    \centering
    \includegraphics[width=3.25in]{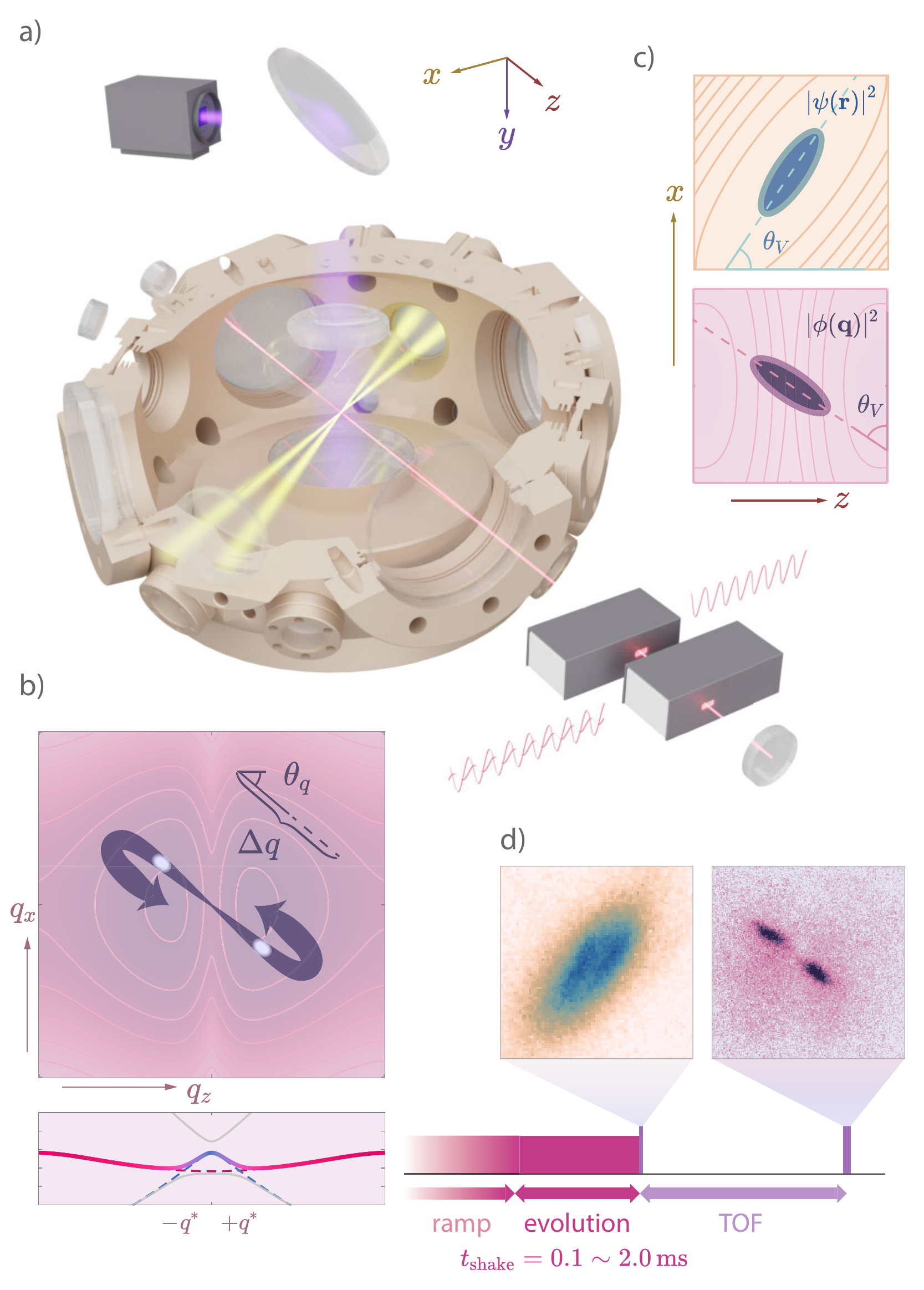}
    \caption{\textbf{a} Experimental setup, showing the lattice (red), dipole trapping (yellow), and imaging (violet) beams. \textbf{b} Effective dispersion in the \(x-z\) plane. The arrows indicate the typical symmetric trajectories of the solitons starting at \(\bm{q} = 0\), characterized by \(\Delta q\) and \(\theta_q\). The line cut below shows the dispersion of quasienergy along the \(z\) axis. The colored solid curve has the double-well feature, and is a hybrid of the ground band (red) and 2nd band (blue) from the non-shaken dispersion (dashed). \textbf{c} Illustration of the initial profile in position and momentum space, with the contour lines depicting the potential and the dispersion close to the zone center. \textbf{d} Typical observations from the experiment: an in-situ image taken at the end of the shaking period, and a time-of-flight (TOF) image, reflecting the momentum space distribution.}
    
    \label{fig:setup}
\end{figure}

We work with interacting molecular BECs of \ce{^6Li_2} in a 3D harmonic trap with a 1D shaken lattice tuned to generate a double-well dispersion. We expect the same physics to occur in atomic BECs, but we use molecular condensates in anticipation of ramping closer to the Feshbach resonance to study strongly interacting Fermi systems. We observe that in 2D momentum space, condensates initially at the unstable Brillouin zone center break apart into two distinct portions (bifurcation), and we categorize the two clusters as solitons given that they each remain concentrated in the momentum density profile during subsequent evolution. We conclude from simulations that this is enabled by 
interaction, and that the trajectory of the solitons can be understood semi-classically as damped movements in two deformed traps (Fig. \ref{fig:setup} b). Our dynamic soliton behavior happens under the condition that the trap's potential energy is comparable to the initial kinetic energy when the BEC is at the dispersion maximum, but not significantly smaller than the interaction energy. Hence our experiment falls in between the large trap limit, which causes chaotic decay, and the no trap limit which yields the static Floquet soliton at the maximum, both of which have been demonstrated previously\cite{mitchell2021floquet}. Compared to experiments that form domains by ramping across the transition and exhibit Kibble-Zurek scaling\cite{clark2016universal,anderson2017direct} our preparation is fast, does not require biasing procedures\cite{yao2022domain} to obtain reproducible results, and works with much weaker overall lattice depth, all of which support applications with more strongly interacting systems where instability and heating will be more prominent.

Our ultracold molecular BEC (mBEC) of \ce{^6Li_2} is loaded in a 1D optical lattice created by a retro-reflected beam of wavelength \(\lambda_\text{L}=\SI{1064}{nm}\), (lattice constant \(a_\text{L} = \SI{532}{nm}\)). We name the lattice direction the \(z\) axis. The lattice's returning beam is diffracted by a pair of acousto-optic modulators (AOMs), each of which is in a double-pass configuration. One of the AOM input signals is modulated by an IQ modulator, where we mix in the shaking signal. We characterize the shaking by the quadrature component's oscillation angular frequency \(\omega\) and its maximum amplitude relative to the static in-phase component \(\xi_\text{max}\). This description is only approximate, however, due to the double-passing of the AOM (see supplemental material). The system can be described by a time-dependent Hamiltonian density
\begin{eqnarray}
    \label{eq:first-principle}
    \mathcal{H} = \Bar{\psi} \bigg[ - \dfrac{\hbar^2\nabla^2}{2m} - \mu \bigg] \psi + \Big[{V}_\text{L}(t) + {V}_\text{trap}\Big]\Bar{\psi}\psi
    + \dfrac{g}{2}(\Bar{\psi}\psi)^2, \nonumber\\
\end{eqnarray}
where \(m\) is the mass of \ce{^6Li_2} molecules, \(\psi\) is the bosonic annihilation operator, \(\mu\) the chemical potential, and \(g\) the interaction strength. The lattice potential contains
\begin{eqnarray}
    V_\text{L}(t) = {V}_\text{L}\qty[\cos(2q_\text{L} z) + \xi_\text{max} \cos(2q_\text{L} z + \dfrac{\pi}{2})\cos(\omega t)]
\end{eqnarray}
\(V_\text{trap}\) represents the trapping potential, \(V_\text{L} = 2.8 E_\text{R}\), where \(E_\text{R}= \frac{h^2}{2m\lambda_\text{L}^2}\) is the recoil energy for \ce{^6Li_2} molecules, \(h\) being the Planck constant and \(q_\text{L} = \frac{2\pi}{\lambda_\text{L}}\) is the lattice light wavevector. The shaking of the lattice allows the first two lattice bands to couple, yielding an effective dispersion relation \(D_{V_\text{L},\omega,\xi_\text{max}}(\bm{q})\), which can be calculated numerically from the shaking parameters \cite{parker2013direct}. The dispersion relation used for the majority of this work is shown in Fig. \ref{fig:setup}.

The details of our apparatus and the Fermi degenerate evaporation process have been described previously \cite{long2018all,long2021spin}. Fig. \ref{fig:setup}a shows the experimental setup. We start with a\ce{^6Li_2} mBEC of approximately 12000 molecules loaded in a harmonic potential formed by both dipole traps and the lattice, which has trapping frequency \SI{810}{Hz} in \(y\), and (210,500) \si{Hz} in the \(x-z\) plane. The lattice shaking is ramped on over \SI{1.2}{ms} (see supplemental material). To prepare most of the BEC in the ground band, the shaking frequency ramps from \SI{80}{kHz} to the target value \(\omega = 2\pi\cross f\), with \(f\) between \SI{45}{kHz} and \SI{72}{kHz}, approximately matching the band gap between the lowest two Bloch bands at the zone center. At this point the condensate fraction is reduced to about \SI{27}{\%}. The lattice shaking is maintained for a period of time \(t_\text{shake}\), before the molecules are released from all traps and lattices. Our absorption imaging system records the optical density integrated along the \(y\) axis, which is perpendicular to the \(z\) axis and parallel to gravity. 

\label{sec:obs}
\begin{figure}
    \centering
    \includegraphics[width=3.25in]{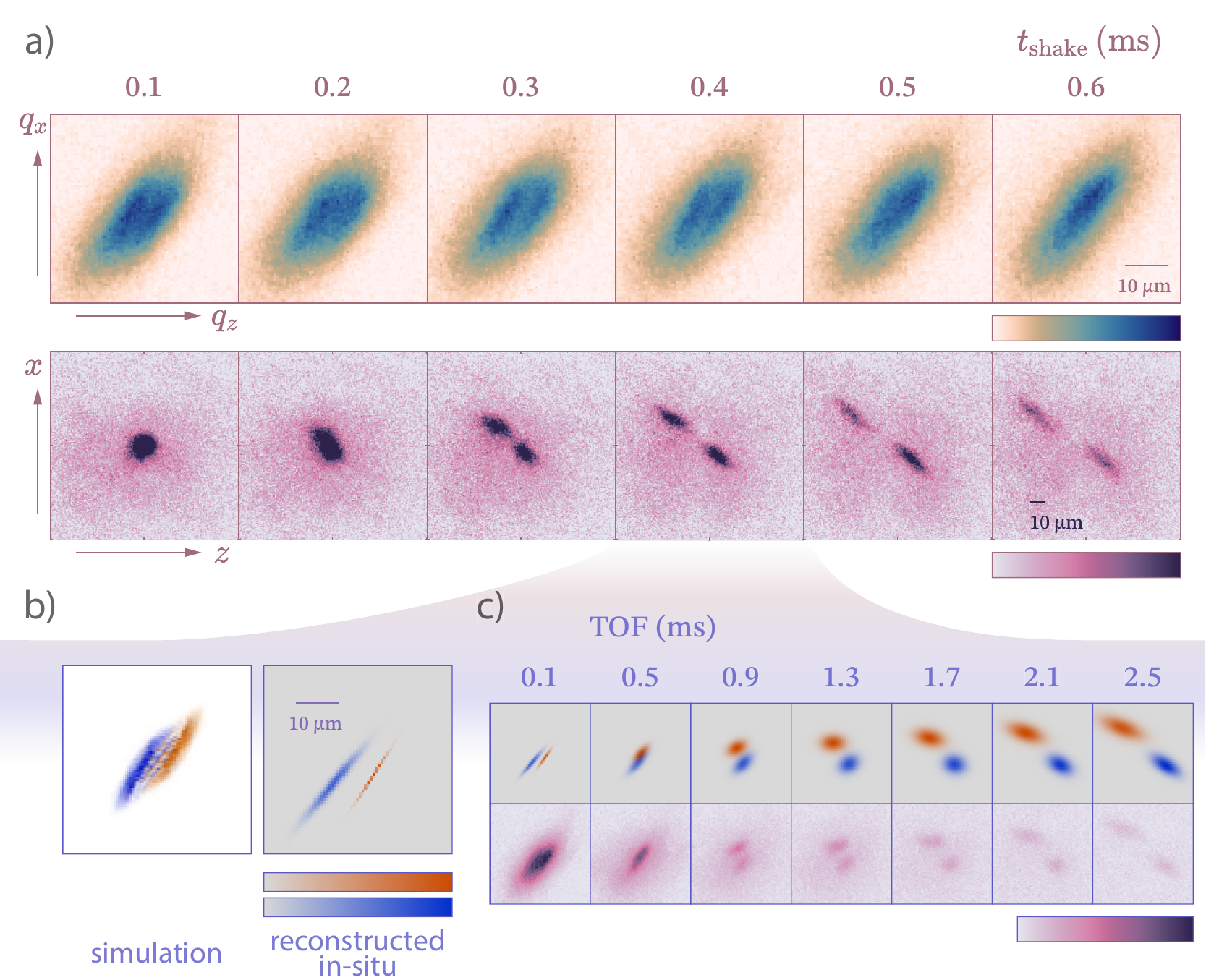}
    \caption{\textbf{a} In-situ (top) and time-of-flight (bottom) images following shaking times \(t_\text{shake}\) from \SI{0.1}{\milli\second} to \SI{0.6}{\milli\second}. Data taken with \(\xi_\text{max} = 0.5\), \(f = \SI{63}{kHz}\), lattice depth \(V_\text{L} = 2.8 E_\text{R}\). The image is cropped to the size of the first Brillouin zone. \textbf{b} colorized simulation (left) and backward-extrapolated in-situ data fitted from (c) (right) showing the component with \(+z\) momentum in red and that with \(-z\) momentum in blue. (c). \textbf{c} Expansion of the mBEC for varying TOF with \(t_\text{shake} = \SI{0.4}{\milli\second}\) (bottom) and their dual-peak Gaussian fits (upper, resp. red and blue). All scale bars are \SI{10}{\micro\meter}.} 
    \label{fig:main-obs}
\end{figure}

When held in the lattice without shaking, the mBEC remains stable for more than \SI{10}{ms}, its spatial profile \(|\psi(\bm{r})|^2\) fitting the contour of the overall potential, which is of an elongated oval shape, with the long axis at an angle with the lattice beams by \(\theta_V=\ang{53}\) (see Fig. \ref{fig:setup}c). The size of the cloud along the \(z\) axis is about \SI{23}{\micro\meter}. If the shaking is turned on, the zone center becomes a saddle point, and the momentum space wavefunction \(\phi(\bm{q})\) lies across both sides of the saddle. Later, the condensate divides into two clusters in momentum space, each half having momentum in opposite directions along an axis close to the direction of strongest confinement in the \(x-z\) plane (see Fig. \ref{fig:main-obs}a). This can be understood by reversing the roles of the effective kinetic energy and the potential energy, where the harmonic trap is seen as an anisotropic parabolic dispersion, and the particles tend to slide down away from the saddle point. The process is availed by a sharp peak in the dispersion profile around \(q_z=0\) (Fig. \ref{fig:setup}b), which results from a low lattice depth, so that most of the cluster initially resides on a linear slope descending to either of the dispersion wells \(\bm{q}^\ast\), in contrast to the stronger lattice case, where the initial cluster \(\phi(\bm{q})\) concentrates on the negative mass region, which may lead to a static soliton \cite{mitchell2021floquet}. The velocity resulting from the different slopes in momentum space create a separation in position space, which is reflected in the observation that the condensate forms a low-density trench in the middle of the sample that can
be resolved in in-situ images, see Fig. \ref{fig:main-obs}b. This can be confirmed by extrapolating the two clusters to their original positions with various time of flights (see Fig. \ref{fig:main-obs}c, supplemental material). Although the gradient of the effective dispersion is along \(z\), during the bifurcation each cluster also acquires momentum in the \(x\) direction, which results from anisotropic effective mass due to the elongated trap. Based on the trap's anisotropy and the angle \(\theta_V\), we can calculate the initial bifurcating angle \(\theta_q|_{t\rightarrow0}\) to be close to \(\SI{30}{\degree}\) (See supplementary materials), which agrees with our measurement.

The two clusters move continuously in momentum space. We characterize the trajectory of the two clusters by their separation \(\Delta q\) and the their relative angle to the \(z\) axis \(\theta_q\) (See Fig. \ref{fig:setup}b). Subsequent evolution shows that the two clusters will each follow a trajectory resembling a damped oscillation around the corresponding dispersion well. In real space the two clusters also collide, corresponding to the vanishing of the trench in the in-situ profiles after \SI{0.5}{\milli\second}. At this point, a density wave could be forming at a wavevector corresponding to the separation \(\Delta q\), whose wavelength \(\sim \SI{1.2}{\micro\meter}\) would be beyond our resolution limit of \(\SI{2}{\micro\meter}\). After collision, we see loss of BEC density from heating and an imbalance of the clusters' molecule number. Although only a fraction of the wavefunction remains condensed, we can still distinguish the two clusters up to \SI{2.0}{\milli\second} of evolution, when the clusters appear to end up around the potential minima \(\pm q^\ast\). This would not happen in the non-interacting case since the effective dispersion well and the trap potential are far from a harmonic condition and would result in much density dispersion (See supplementary materials Fig. S4), but the repulsive interaction enables the formation of 2D solitons in momentum space. 

\begin{figure*}
    \centering
    \includegraphics[width=\linewidth]{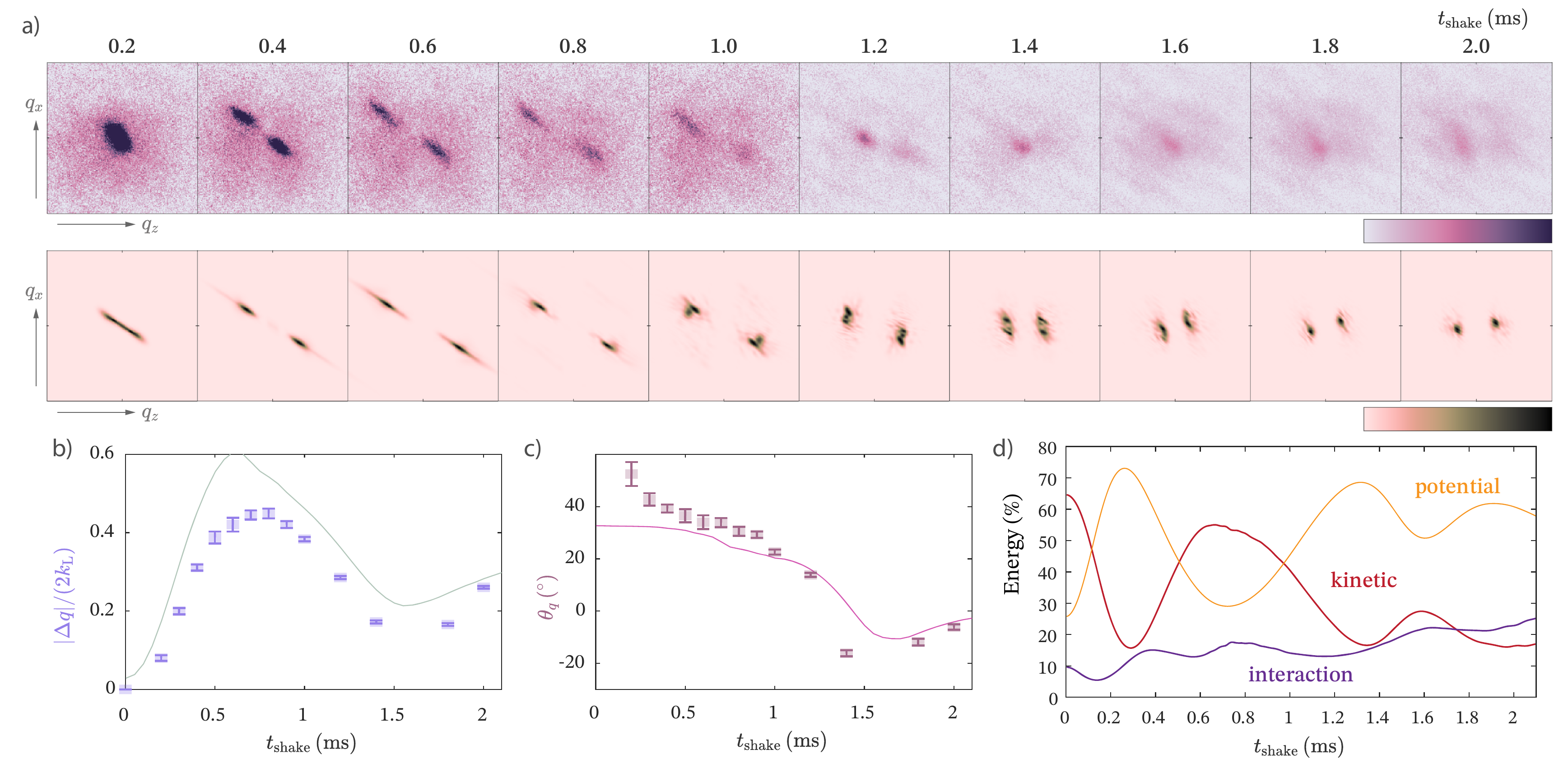}
    \caption{ \textbf{a} Evolution of the momentum distribution, the top row is obtained by TOF imaging, for shaking times \(t_\text{shake}=\SI{0.2}{\milli\second}\sim\SI{2.0}{\milli\second}\), the bottom row is from simulation under similar conditions. \textbf{b} Measurement of the peak separation \(\Delta q\) between the two clusters  (points with error bars), together with the corresponding value from GP simulations (solid lines). The lines correspond to results of the simulation with a range of initial sizes. \textbf{c} Measurement and GP simulation results for the angle of separation, \(\theta_q\). \textbf{d} Evolution of the kinetic energy (red), trap potential energy (yellow), and interaction energy (violet) from one of the simulations.}
    \label{fig:trajectory}
\end{figure*}

We performed numeric simulation using a model with an effective dispersion \(\hat D_{V_\text{L},\omega,\xi_\text{max}}(\bm{q})\) (see Fig. \ref{fig:trajectory}). The evolution of the BEC wavefunction \(\psi(\bm{r})\) can be described by the non-linear Gross-Pitaevskii equation derived from (\ref{eq:first-principle}):
\begin{equation}
    \label{eq:gpe}
     i \hbar \pdv{\psi}{t} = \bigg\{ - \dfrac{\hbar^2\bm{\nabla}^2}{2 m} + V_\text{trap}(\bm{r})) + g\qty|\psi^2|  + V_\text{L}(t) \bigg\} \psi.
\end{equation}
The trapping potential \(V_\text{trap}\) for the simulation is adjusted to match the timescale of the experiment, which corresponds with a 25\% reduction of the trap frequency. The Hamiltonian is periodic in time with period \(T = \frac{2\pi}{\omega}\). Floquet theory tells us that the solution to equation (\ref{eq:gpe}) will be in the form \(\ket{\psi(t)} = \sum_n a_n \ket{u_n(t)} e^{-i\epsilon_n t/\hbar}\), where \(\ket{u_n(t)}\) are the fast-changing Floquet modes. Each mode is \(t\)-dependent and periodic in \(T\). The method separates the evolution into a fast-repeating micro-motion \(\ket{u_n(t)}\) and the slow-evolving dynamic \(\epsilon_n\). The latter of these is where our interest mainly lies, which is described by the effective Hamiltonian (\ref{eq:floq-hamilton}).
\begin{eqnarray}  
    \label{eq:floq-hamilton} 
    \hat H_{\tau_0, \text{Floq}} &\approx & 
    \dfrac{i\hbar}{T} \log \prod_{\tau_0}^{\tau_0 + T} \exp{-\dfrac{i}{\hbar}\bigg[\hat K + \hat V_0 + \hat V_{\frac{\pi}{2}} \xi(t)\bigg] \dd t}  \nonumber\\
    && + \hat{V}_\text{trap} + g\qty|\psi^2| \label{eq:dspseff1}\\
    \label{eq:dspseff2}
    &\approx & \hat D_{V_\text{L},\omega,\xi_\text{max}} + \hat{V}_\text{trap} + g\qty|\psi^2|
\end{eqnarray}

To isolate the effective dispersion operator \(\hat D_{V_\text{L},\omega,\xi_\text{max}}\), we take advantage of the fact that the fast changing term in \(\hat H\) will couple mostly with the kinetic term. In (\ref{eq:dspseff1}), \(\hat K\) is the kinetic energy operator, \(\hat V_0\) and \(\hat V_{\frac{\pi}{2}}\) are lattice operators for lattices with the same depth \(V_\text{L}\) at phase \(0\) and \(\frac{\pi}{2}\), representing the static lattice and shaken lattice respectively. \(\hat D_{V_\text{L},\omega,\xi_\text{max}}\) is then calculated numerically and results in the double-well form.  \\

\begin{figure}
    \centering
    \includegraphics[width=3.25in]{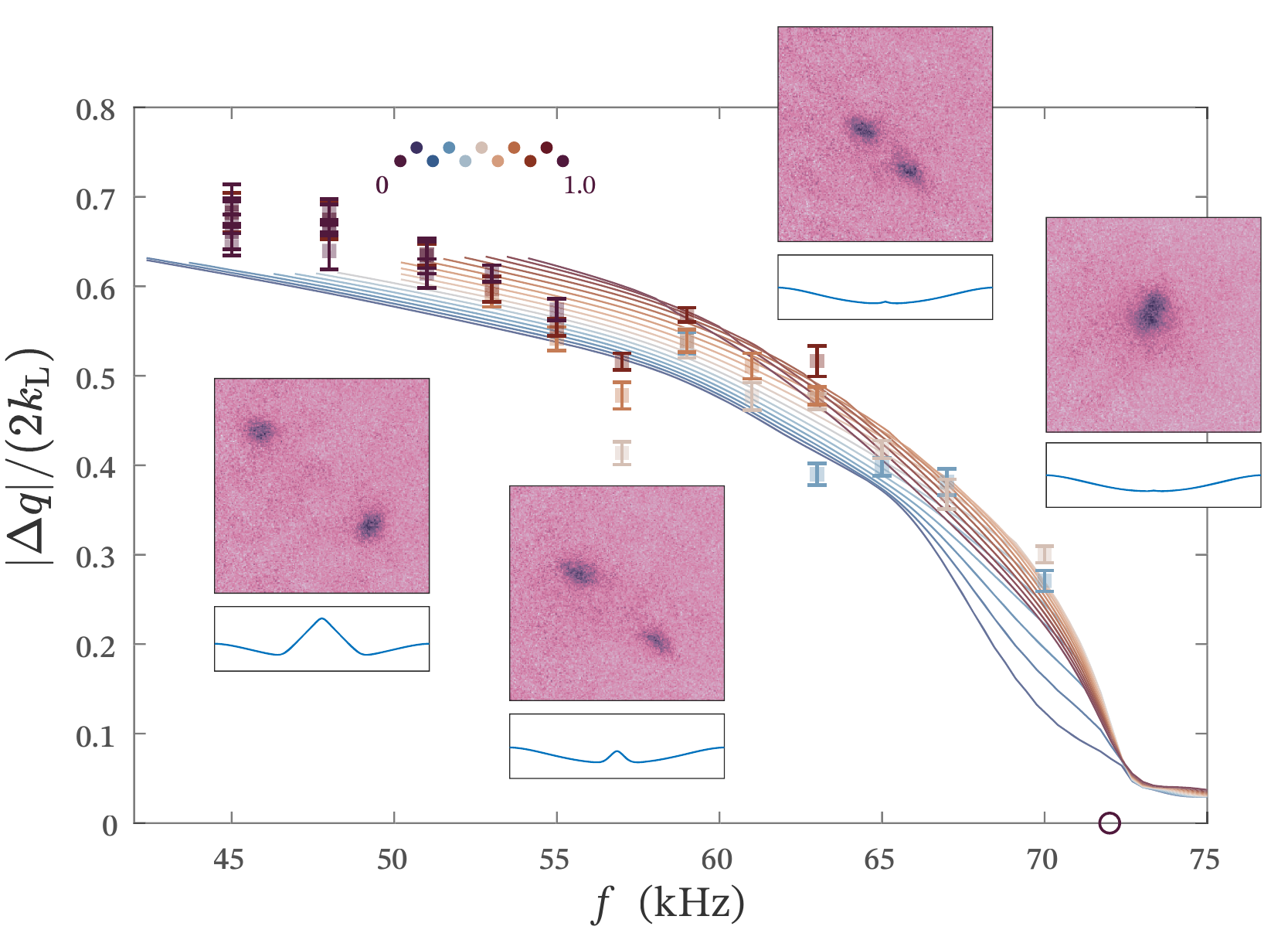} 
    \caption{ Variation of the peak separation \(\Delta q\) at \(t_\text{shake} = \SI{0.4}{\milli\second}\) for a range of shaking frequencies from \(f = 45 \sim 72 \,\si{kHz}\) (data points with error bars) together with results from the GP simulation (solid lines). The inset images are TOF momentum distributions from which \(\Delta q\) is determined, together with the calculated dispersion appropriate to that frequency (from left to right: \(48,\,63,\,70,\,72\si{kHz}\)). The color of the lines and data points indicate different shaken amplitudes \(\xi_\text{max}\). }
    \label{fig:freqvar}
\end{figure}

Our simulation starts at the approximate ground state with \(V_\text{L} = 0\) with a cluster size of \(\SI{16}{\micro\meter}\), which is empirically chosen slightly smaller than the experimental size to account for the 27\% condensate fraction. We then take a Trotter product of \(\exp{-(i + \Gamma)\hat D \Delta t/\hbar}\) and \(\exp{-(i + \Gamma) (\hat V_\text{trap} + g\qty|\psi^2|) \Delta t/\hbar}\) by calculating their matrix representations in momentum space and position space respectively (see supplemental material), where \(\Gamma\) is an empirical dissipation coefficient \cite{ranccon2014equilibrating}. Unfortunately, the single-band GP method fails to capture the interaction that exists between higher band states and the ground band, that is it assumes the ground-to-first excited transition frequency is not affected by interactions. In reality this frequency, or equivalently the phase velocity of acoustic waves with the period of the lattice spacing, is shifted for strongly interacting systems. To correct for this, we introduce a modified recoil energy \(E_\text{R}'\) acquired from the measured phase velocity in a Kapitsa-Dirac experiment, which is about \SI{10}{\%} larger than the \(E_\text{R}\) calculated from the molecular mass. This is consistent with the observation that the shaking frequency threshold for creating a momentum bifurcation is generally larger than the free particle band gap would allow (See Fig. \ref{fig:freqvar}). The experiment and simulation under the same nominal conditions shows the same trend of bifurcation and a similar subsequent trajectory for the pair of solitons (Fig. \ref{fig:trajectory}). The simulation allows us to tune a wider range of parameters. We discovered that other than the sharp peak feature in the dispersion relation, the clear bifurcation requires the initial potential energy to be smaller than the total kinetic energy, which allows for the following periodic conversion between the two, corresponding to the oscillation in the trajectory. Our result can serve as a cross-over region between the two dynamical phases of chaotic decay at large confinement and static soliton at no confinement \cite{mitchell2021floquet} (See Fig. S5 in supplementary materials). 

By tuning the shaking frequency \(\omega\) we effectively change how much the second band protrudes into the ground band in the Floquet picture, and the separation between the quasienergy minima \(2\abs{q^\ast}\). At lower frequencies, the momentum space clusters glide along a longer slope and exhibits larger separation \(\Delta q\) at the same \(t_\text{shake}\). At higher frequencies, the target band is not hybridized and \(q^\ast=0\), the bifurcation will thus not happen. Figure \ref{fig:freqvar} displays the evolution at a range of \(f\), which matches our simulation results. Increasing the amplitude of the shaking does not change the band shape significantly for the conditions of our experiment, and therefore does not affect the simulation results.


In conclusion, we have successfully loaded a strongly interacting \ce{{}^6Li_2} molecular BEC into a shaken lattice, with a Floquet engineered double-well dispersion. The resulting dynamics differs significantly from previous observations, featuring a bifurcation into two clusters in momentum space and a trench in the position space density. We can describe the dynamics semi-classically by mapping it to a model featuring anisotropic mass and reproduce it qualitatively with a Gross-Pitaevskii simulation. We show that this phenomenon is enabled by the repulsive interaction and occurs over a substantial range of dispersion relations tuned via the lattice shaking frequency. The rapid splitting into two clusters suggests a novel method for preparing molecular-Bose or Fermi systems in non-equilibrium states with exotic engineered dispersions. For example, if after the initial splitting, conditions could be altered by the appropriate counter-diabatic protocol, one might produce a ``soft-landing'' for the two clusters leaving them as quasi-stable domains. In such a way even strongly interacting systems, such as unitary Fermi gases, might be prepared in domain configurations that couldn't be achieved adiabatically due to unwanted collisional heating \cite{choudhury2014stability, choudhury2015stability, choudhury2015transverse}.


\begin{acknowledgments}
We acknowledge funding from NSF CAREER award No. 1941985. We also thank Carlos Sa de Melo for comments on the manuscript.

\end{acknowledgments}

\appendix

%


\begin{thebibliography}{33}%
\makeatletter
\providecommand \@ifxundefined [1]{%
 \@ifx{#1\undefined}
}%
\providecommand \@ifnum [1]{%
 \ifnum #1\expandafter \@firstoftwo
 \else \expandafter \@secondoftwo
 \fi
}%
\providecommand \@ifx [1]{%
 \ifx #1\expandafter \@firstoftwo
 \else \expandafter \@secondoftwo
 \fi
}%
\providecommand \natexlab [1]{#1}%
\providecommand \enquote  [1]{``#1''}%
\providecommand \bibnamefont  [1]{#1}%
\providecommand \bibfnamefont [1]{#1}%
\providecommand \citenamefont [1]{#1}%
\providecommand \href@noop [0]{\@secondoftwo}%
\providecommand \href [0]{\begingroup \@sanitize@url \@href}%
\providecommand \@href[1]{\@@startlink{#1}\@@href}%
\providecommand \@@href[1]{\endgroup#1\@@endlink}%
\providecommand \@sanitize@url [0]{\catcode `\\12\catcode `\$12\catcode
  `\&12\catcode `\#12\catcode `\^12\catcode `\_12\catcode `\%12\relax}%
\providecommand \@@startlink[1]{}%
\providecommand \@@endlink[0]{}%
\providecommand \url  [0]{\begingroup\@sanitize@url \@url }%
\providecommand \@url [1]{\endgroup\@href {#1}{\urlprefix }}%
\providecommand \urlprefix  [0]{URL }%
\providecommand \Eprint [0]{\href }%
\providecommand \doibase [0]{https://doi.org/}%
\providecommand \selectlanguage [0]{\@gobble}%
\providecommand \bibinfo  [0]{\@secondoftwo}%
\providecommand \bibfield  [0]{\@secondoftwo}%
\providecommand \translation [1]{[#1]}%
\providecommand \BibitemOpen [0]{}%
\providecommand \bibitemStop [0]{}%
\providecommand \bibitemNoStop [0]{.\EOS\space}%
\providecommand \EOS [0]{\spacefactor3000\relax}%
\providecommand \BibitemShut  [1]{\csname bibitem#1\endcsname}%
\let\auto@bib@innerbib\@empty
\bibitem [{\citenamefont {Mitchell}\ \emph {et~al.}(2021)\citenamefont
  {Mitchell}, \citenamefont {Di~Carli}, \citenamefont {Sinuco-Le{\'o}n},
  \citenamefont {La~Rooij}, \citenamefont {Kuhr},\ and\ \citenamefont
  {Haller}}]{mitchell2021floquet}%
  \BibitemOpen
  \bibfield  {author} {\bibinfo {author} {\bibfnamefont {M.}~\bibnamefont
  {Mitchell}}, \bibinfo {author} {\bibfnamefont {A.}~\bibnamefont {Di~Carli}},
  \bibinfo {author} {\bibfnamefont {G.}~\bibnamefont {Sinuco-Le{\'o}n}},
  \bibinfo {author} {\bibfnamefont {A.}~\bibnamefont {La~Rooij}}, \bibinfo
  {author} {\bibfnamefont {S.}~\bibnamefont {Kuhr}},\ and\ \bibinfo {author}
  {\bibfnamefont {E.}~\bibnamefont {Haller}},\ }\bibfield  {title} {\bibinfo
  {title} {Floquet solitons and dynamics of periodically driven matter waves
  with negative effective mass},\ }\href@noop {} {\bibfield  {journal}
  {\bibinfo  {journal} {Physical Review Letters}\ }\textbf {\bibinfo {volume}
  {127}},\ \bibinfo {pages} {243603} (\bibinfo {year} {2021})}\BibitemShut
  {NoStop}%
\bibitem [{\citenamefont {Serbyn}\ \emph {et~al.}(2021)\citenamefont {Serbyn},
  \citenamefont {Abanin},\ and\ \citenamefont {Papi{\'c}}}]{serbyn2021quantum}%
  \BibitemOpen
  \bibfield  {author} {\bibinfo {author} {\bibfnamefont {M.}~\bibnamefont
  {Serbyn}}, \bibinfo {author} {\bibfnamefont {D.~A.}\ \bibnamefont {Abanin}},\
  and\ \bibinfo {author} {\bibfnamefont {Z.}~\bibnamefont {Papi{\'c}}},\
  }\bibfield  {title} {\bibinfo {title} {Quantum many-body scars and weak
  breaking of ergodicity},\ }\href@noop {} {\bibfield  {journal} {\bibinfo
  {journal} {Nature Physics}\ }\textbf {\bibinfo {volume} {17}},\ \bibinfo
  {pages} {675} (\bibinfo {year} {2021})}\BibitemShut {NoStop}%
\bibitem [{\citenamefont {Jepsen}\ \emph {et~al.}(2022)\citenamefont {Jepsen},
  \citenamefont {Lee}, \citenamefont {Lin}, \citenamefont {Dimitrova},
  \citenamefont {Margalit}, \citenamefont {Ho},\ and\ \citenamefont
  {Ketterle}}]{jepsen2022long}%
  \BibitemOpen
  \bibfield  {author} {\bibinfo {author} {\bibfnamefont {P.~N.}\ \bibnamefont
  {Jepsen}}, \bibinfo {author} {\bibfnamefont {Y.~K.~E.}\ \bibnamefont {Lee}},
  \bibinfo {author} {\bibfnamefont {H.}~\bibnamefont {Lin}}, \bibinfo {author}
  {\bibfnamefont {I.}~\bibnamefont {Dimitrova}}, \bibinfo {author}
  {\bibfnamefont {Y.}~\bibnamefont {Margalit}}, \bibinfo {author}
  {\bibfnamefont {W.~W.}\ \bibnamefont {Ho}},\ and\ \bibinfo {author}
  {\bibfnamefont {W.}~\bibnamefont {Ketterle}},\ }\bibfield  {title} {\bibinfo
  {title} {Long-lived phantom helix states in heisenberg quantum magnets},\
  }\href@noop {} {\bibfield  {journal} {\bibinfo  {journal} {Nature Physics}\
  }\textbf {\bibinfo {volume} {18}},\ \bibinfo {pages} {899} (\bibinfo {year}
  {2022})}\BibitemShut {NoStop}%
\bibitem [{\citenamefont {Lahaye}\ \emph {et~al.}(2008)\citenamefont {Lahaye},
  \citenamefont {Metz}, \citenamefont {Froehlich}, \citenamefont {Koch},
  \citenamefont {Meister}, \citenamefont {Griesmaier}, \citenamefont {Pfau},
  \citenamefont {Saito}, \citenamefont {Kawaguchi},\ and\ \citenamefont
  {Ueda}}]{lahaye2008d}%
  \BibitemOpen
  \bibfield  {author} {\bibinfo {author} {\bibfnamefont {T.}~\bibnamefont
  {Lahaye}}, \bibinfo {author} {\bibfnamefont {J.}~\bibnamefont {Metz}},
  \bibinfo {author} {\bibfnamefont {B.}~\bibnamefont {Froehlich}}, \bibinfo
  {author} {\bibfnamefont {T.}~\bibnamefont {Koch}}, \bibinfo {author}
  {\bibfnamefont {M.}~\bibnamefont {Meister}}, \bibinfo {author} {\bibfnamefont
  {A.}~\bibnamefont {Griesmaier}}, \bibinfo {author} {\bibfnamefont
  {T.}~\bibnamefont {Pfau}}, \bibinfo {author} {\bibfnamefont {H.}~\bibnamefont
  {Saito}}, \bibinfo {author} {\bibfnamefont {Y.}~\bibnamefont {Kawaguchi}},\
  and\ \bibinfo {author} {\bibfnamefont {M.}~\bibnamefont {Ueda}},\ }\bibfield
  {title} {\bibinfo {title} {d-wave collapse and explosion of a dipolar
  bose-einstein condensate},\ }\href@noop {} {\bibfield  {journal} {\bibinfo
  {journal} {Physical review letters}\ }\textbf {\bibinfo {volume} {101}},\
  \bibinfo {pages} {080401} (\bibinfo {year} {2008})}\BibitemShut {NoStop}%
\bibitem [{\citenamefont {Clark}\ \emph {et~al.}(2017)\citenamefont {Clark},
  \citenamefont {Gaj}, \citenamefont {Feng},\ and\ \citenamefont
  {Chin}}]{clark2017collective}%
  \BibitemOpen
  \bibfield  {author} {\bibinfo {author} {\bibfnamefont {L.~W.}\ \bibnamefont
  {Clark}}, \bibinfo {author} {\bibfnamefont {A.}~\bibnamefont {Gaj}}, \bibinfo
  {author} {\bibfnamefont {L.}~\bibnamefont {Feng}},\ and\ \bibinfo {author}
  {\bibfnamefont {C.}~\bibnamefont {Chin}},\ }\bibfield  {title} {\bibinfo
  {title} {Collective emission of matter-wave jets from driven bose--einstein
  condensates},\ }\href@noop {} {\bibfield  {journal} {\bibinfo  {journal}
  {Nature}\ }\textbf {\bibinfo {volume} {551}},\ \bibinfo {pages} {356}
  (\bibinfo {year} {2017})}\BibitemShut {NoStop}%
\bibitem [{\citenamefont {Clark}\ \emph {et~al.}(2016)\citenamefont {Clark},
  \citenamefont {Feng},\ and\ \citenamefont {Chin}}]{clark2016universal}%
  \BibitemOpen
  \bibfield  {author} {\bibinfo {author} {\bibfnamefont {L.~W.}\ \bibnamefont
  {Clark}}, \bibinfo {author} {\bibfnamefont {L.}~\bibnamefont {Feng}},\ and\
  \bibinfo {author} {\bibfnamefont {C.}~\bibnamefont {Chin}},\ }\bibfield
  {title} {\bibinfo {title} {Universal space-time scaling symmetry in the
  dynamics of bosons across a quantum phase transition},\ }\href@noop {}
  {\bibfield  {journal} {\bibinfo  {journal} {Science}\ }\textbf {\bibinfo
  {volume} {354}},\ \bibinfo {pages} {606} (\bibinfo {year}
  {2016})}\BibitemShut {NoStop}%
\bibitem [{\citenamefont {Yao}\ \emph {et~al.}(2022)\citenamefont {Yao},
  \citenamefont {Zhang},\ and\ \citenamefont {Chin}}]{yao2022domain}%
  \BibitemOpen
  \bibfield  {author} {\bibinfo {author} {\bibfnamefont {K.-X.}\ \bibnamefont
  {Yao}}, \bibinfo {author} {\bibfnamefont {Z.}~\bibnamefont {Zhang}},\ and\
  \bibinfo {author} {\bibfnamefont {C.}~\bibnamefont {Chin}},\ }\bibfield
  {title} {\bibinfo {title} {Domain-wall dynamics in bose--einstein condensates
  with synthetic gauge fields},\ }\href@noop {} {\bibfield  {journal} {\bibinfo
   {journal} {Nature}\ }\textbf {\bibinfo {volume} {602}},\ \bibinfo {pages}
  {68} (\bibinfo {year} {2022})}\BibitemShut {NoStop}%
\bibitem [{\citenamefont {Sels}\ and\ \citenamefont
  {Polkovnikov}(2017)}]{sels_minimizing_2017}%
  \BibitemOpen
  \bibfield  {author} {\bibinfo {author} {\bibfnamefont {D.}~\bibnamefont
  {Sels}}\ and\ \bibinfo {author} {\bibfnamefont {A.}~\bibnamefont
  {Polkovnikov}},\ }\bibfield  {title} {\bibinfo {title} {Minimizing
  irreversible losses in quantum systems by local counterdiabatic driving},\
  }\href {https://doi.org/10.1073/pnas.1619826114} {\bibfield  {journal}
  {\bibinfo  {journal} {Proceedings of the National Academy of Sciences}\
  }\textbf {\bibinfo {volume} {114}},\ \bibinfo {pages} {E3909} (\bibinfo
  {year} {2017})}\BibitemShut {NoStop}%
\bibitem [{\citenamefont {Eckardt}(2017)}]{eckardt2017colloquium}%
  \BibitemOpen
  \bibfield  {author} {\bibinfo {author} {\bibfnamefont {A.}~\bibnamefont
  {Eckardt}},\ }\bibfield  {title} {\bibinfo {title} {Colloquium: Atomic
  quantum gases in periodically driven optical lattices},\ }\href@noop {}
  {\bibfield  {journal} {\bibinfo  {journal} {Reviews of Modern Physics}\
  }\textbf {\bibinfo {volume} {89}},\ \bibinfo {pages} {011004} (\bibinfo
  {year} {2017})}\BibitemShut {NoStop}%
\bibitem [{\citenamefont {Weitenberg}\ and\ \citenamefont
  {Simonet}(2021)}]{weitenberg2021tailoring}%
  \BibitemOpen
  \bibfield  {author} {\bibinfo {author} {\bibfnamefont {C.}~\bibnamefont
  {Weitenberg}}\ and\ \bibinfo {author} {\bibfnamefont {J.}~\bibnamefont
  {Simonet}},\ }\bibfield  {title} {\bibinfo {title} {Tailoring quantum gases
  by floquet engineering},\ }\href@noop {} {\bibfield  {journal} {\bibinfo
  {journal} {Nature Physics}\ }\textbf {\bibinfo {volume} {17}},\ \bibinfo
  {pages} {1342} (\bibinfo {year} {2021})}\BibitemShut {NoStop}%
\bibitem [{\citenamefont {Zahn}\ \emph {et~al.}(2022)\citenamefont {Zahn},
  \citenamefont {Singh}, \citenamefont {Kosch}, \citenamefont {Asteria},
  \citenamefont {Freystatzky}, \citenamefont {Sengstock}, \citenamefont
  {Mathey},\ and\ \citenamefont {Weitenberg}}]{zahn2022formation}%
  \BibitemOpen
  \bibfield  {author} {\bibinfo {author} {\bibfnamefont {H.~P.}\ \bibnamefont
  {Zahn}}, \bibinfo {author} {\bibfnamefont {V.~P.}\ \bibnamefont {Singh}},
  \bibinfo {author} {\bibfnamefont {M.~N.}\ \bibnamefont {Kosch}}, \bibinfo
  {author} {\bibfnamefont {L.}~\bibnamefont {Asteria}}, \bibinfo {author}
  {\bibfnamefont {L.}~\bibnamefont {Freystatzky}}, \bibinfo {author}
  {\bibfnamefont {K.}~\bibnamefont {Sengstock}}, \bibinfo {author}
  {\bibfnamefont {L.}~\bibnamefont {Mathey}},\ and\ \bibinfo {author}
  {\bibfnamefont {C.}~\bibnamefont {Weitenberg}},\ }\bibfield  {title}
  {\bibinfo {title} {Formation of spontaneous density-wave patterns in dc
  driven lattices},\ }\href {https://doi.org/10.1103/PhysRevX.12.021014}
  {\bibfield  {journal} {\bibinfo  {journal} {Phys. Rev. X}\ }\textbf {\bibinfo
  {volume} {12}},\ \bibinfo {pages} {021014} (\bibinfo {year}
  {2022})}\BibitemShut {NoStop}%
\bibitem [{\citenamefont {Struck}\ \emph {et~al.}(2014)\citenamefont {Struck},
  \citenamefont {Simonet},\ and\ \citenamefont {Sengstock}}]{struck2014spin}%
  \BibitemOpen
  \bibfield  {author} {\bibinfo {author} {\bibfnamefont {J.}~\bibnamefont
  {Struck}}, \bibinfo {author} {\bibfnamefont {J.}~\bibnamefont {Simonet}},\
  and\ \bibinfo {author} {\bibfnamefont {K.}~\bibnamefont {Sengstock}},\
  }\bibfield  {title} {\bibinfo {title} {Spin-orbit coupling in periodically
  driven optical lattices},\ }\href
  {https://doi.org/10.1103/PhysRevA.90.031601} {\bibfield  {journal} {\bibinfo
  {journal} {Phys. Rev. A}\ }\textbf {\bibinfo {volume} {90}},\ \bibinfo
  {pages} {031601} (\bibinfo {year} {2014})}\BibitemShut {NoStop}%
\bibitem [{\citenamefont {Struck}\ \emph {et~al.}(2012)\citenamefont {Struck},
  \citenamefont {{\"O}lschl{\"a}ger}, \citenamefont {Weinberg}, \citenamefont
  {Hauke}, \citenamefont {Simonet}, \citenamefont {Eckardt}, \citenamefont
  {Lewenstein}, \citenamefont {Sengstock},\ and\ \citenamefont
  {Windpassinger}}]{struck2012tunable}%
  \BibitemOpen
  \bibfield  {author} {\bibinfo {author} {\bibfnamefont {J.}~\bibnamefont
  {Struck}}, \bibinfo {author} {\bibfnamefont {C.}~\bibnamefont
  {{\"O}lschl{\"a}ger}}, \bibinfo {author} {\bibfnamefont {M.}~\bibnamefont
  {Weinberg}}, \bibinfo {author} {\bibfnamefont {P.}~\bibnamefont {Hauke}},
  \bibinfo {author} {\bibfnamefont {J.}~\bibnamefont {Simonet}}, \bibinfo
  {author} {\bibfnamefont {A.}~\bibnamefont {Eckardt}}, \bibinfo {author}
  {\bibfnamefont {M.}~\bibnamefont {Lewenstein}}, \bibinfo {author}
  {\bibfnamefont {K.}~\bibnamefont {Sengstock}},\ and\ \bibinfo {author}
  {\bibfnamefont {P.}~\bibnamefont {Windpassinger}},\ }\bibfield  {title}
  {\bibinfo {title} {Tunable gauge potential for neutral and spinless particles
  in driven optical lattices},\ }\href@noop {} {\bibfield  {journal} {\bibinfo
  {journal} {Physical review letters}\ }\textbf {\bibinfo {volume} {108}},\
  \bibinfo {pages} {225304} (\bibinfo {year} {2012})}\BibitemShut {NoStop}%
\bibitem [{\citenamefont {Jotzu}\ \emph {et~al.}(2014)\citenamefont {Jotzu},
  \citenamefont {Messer}, \citenamefont {Desbuquois}, \citenamefont {Lebrat},
  \citenamefont {Uehlinger}, \citenamefont {Greif},\ and\ \citenamefont
  {Esslinger}}]{jotzu2014experimental}%
  \BibitemOpen
  \bibfield  {author} {\bibinfo {author} {\bibfnamefont {G.}~\bibnamefont
  {Jotzu}}, \bibinfo {author} {\bibfnamefont {M.}~\bibnamefont {Messer}},
  \bibinfo {author} {\bibfnamefont {R.}~\bibnamefont {Desbuquois}}, \bibinfo
  {author} {\bibfnamefont {M.}~\bibnamefont {Lebrat}}, \bibinfo {author}
  {\bibfnamefont {T.}~\bibnamefont {Uehlinger}}, \bibinfo {author}
  {\bibfnamefont {D.}~\bibnamefont {Greif}},\ and\ \bibinfo {author}
  {\bibfnamefont {T.}~\bibnamefont {Esslinger}},\ }\bibfield  {title} {\bibinfo
  {title} {Experimental realization of the topological haldane model with
  ultracold fermions},\ }\href@noop {} {\bibfield  {journal} {\bibinfo
  {journal} {Nature}\ }\textbf {\bibinfo {volume} {515}},\ \bibinfo {pages}
  {237} (\bibinfo {year} {2014})}\BibitemShut {NoStop}%
\bibitem [{\citenamefont {Sandholzer}(2022)}]{sandholzer2022floquet}%
  \BibitemOpen
  \bibfield  {author} {\bibinfo {author} {\bibfnamefont {K.}~\bibnamefont
  {Sandholzer}},\ }\emph {\bibinfo {title} {Floquet engineering of ultracold
  atoms in optical lattices}},\ \href@noop {} {Ph.D. thesis},\ \bibinfo
  {school} {ETH Zurich} (\bibinfo {year} {2022})\BibitemShut {NoStop}%
\bibitem [{\citenamefont {Parker}\ \emph {et~al.}(2013)\citenamefont {Parker},
  \citenamefont {Ha},\ and\ \citenamefont {Chin}}]{parker2013direct}%
  \BibitemOpen
  \bibfield  {author} {\bibinfo {author} {\bibfnamefont {C.~V.}\ \bibnamefont
  {Parker}}, \bibinfo {author} {\bibfnamefont {L.-C.}\ \bibnamefont {Ha}},\
  and\ \bibinfo {author} {\bibfnamefont {C.}~\bibnamefont {Chin}},\ }\bibfield
  {title} {\bibinfo {title} {Direct observation of effective ferromagnetic
  domains of cold atoms in a shaken optical lattice},\ }\href@noop {}
  {\bibfield  {journal} {\bibinfo  {journal} {Nature Physics}\ }\textbf
  {\bibinfo {volume} {9}},\ \bibinfo {pages} {769} (\bibinfo {year}
  {2013})}\BibitemShut {NoStop}%
\bibitem [{\citenamefont {Anderson}\ \emph {et~al.}(2017)\citenamefont
  {Anderson}, \citenamefont {Clark}, \citenamefont {Crawford}, \citenamefont
  {Glatz}, \citenamefont {Aranson}, \citenamefont {Scherpelz}, \citenamefont
  {Feng}, \citenamefont {Chin},\ and\ \citenamefont
  {Levin}}]{anderson2017direct}%
  \BibitemOpen
  \bibfield  {author} {\bibinfo {author} {\bibfnamefont {B.~M.}\ \bibnamefont
  {Anderson}}, \bibinfo {author} {\bibfnamefont {L.~W.}\ \bibnamefont {Clark}},
  \bibinfo {author} {\bibfnamefont {J.}~\bibnamefont {Crawford}}, \bibinfo
  {author} {\bibfnamefont {A.}~\bibnamefont {Glatz}}, \bibinfo {author}
  {\bibfnamefont {I.~S.}\ \bibnamefont {Aranson}}, \bibinfo {author}
  {\bibfnamefont {P.}~\bibnamefont {Scherpelz}}, \bibinfo {author}
  {\bibfnamefont {L.}~\bibnamefont {Feng}}, \bibinfo {author} {\bibfnamefont
  {C.}~\bibnamefont {Chin}},\ and\ \bibinfo {author} {\bibfnamefont
  {K.}~\bibnamefont {Levin}},\ }\bibfield  {title} {\bibinfo {title} {Direct
  lattice shaking of bose condensates: Finite momentum superfluids},\
  }\href@noop {} {\bibfield  {journal} {\bibinfo  {journal} {Physical review
  letters}\ }\textbf {\bibinfo {volume} {118}},\ \bibinfo {pages} {220401}
  (\bibinfo {year} {2017})}\BibitemShut {NoStop}%
\bibitem [{\citenamefont {Song}\ \emph {et~al.}(2022)\citenamefont {Song},
  \citenamefont {Dutta}, \citenamefont {Bhave}, \citenamefont {Yu},
  \citenamefont {Carter}, \citenamefont {Cooper},\ and\ \citenamefont
  {Schneider}}]{song2022realizing}%
  \BibitemOpen
  \bibfield  {author} {\bibinfo {author} {\bibfnamefont {B.}~\bibnamefont
  {Song}}, \bibinfo {author} {\bibfnamefont {S.}~\bibnamefont {Dutta}},
  \bibinfo {author} {\bibfnamefont {S.}~\bibnamefont {Bhave}}, \bibinfo
  {author} {\bibfnamefont {J.-C.}\ \bibnamefont {Yu}}, \bibinfo {author}
  {\bibfnamefont {E.}~\bibnamefont {Carter}}, \bibinfo {author} {\bibfnamefont
  {N.}~\bibnamefont {Cooper}},\ and\ \bibinfo {author} {\bibfnamefont
  {U.}~\bibnamefont {Schneider}},\ }\bibfield  {title} {\bibinfo {title}
  {Realizing discontinuous quantum phase transitions in a strongly correlated
  driven optical lattice},\ }\href@noop {} {\bibfield  {journal} {\bibinfo
  {journal} {Nature Physics}\ }\textbf {\bibinfo {volume} {18}},\ \bibinfo
  {pages} {259} (\bibinfo {year} {2022})}\BibitemShut {NoStop}%
\bibitem [{\citenamefont {Kele{\c{s}}}\ \emph {et~al.}(2017)\citenamefont
  {Kele{\c{s}}}, \citenamefont {Zhao},\ and\ \citenamefont
  {Liu}}]{kelecs2017effective}%
  \BibitemOpen
  \bibfield  {author} {\bibinfo {author} {\bibfnamefont {A.}~\bibnamefont
  {Kele{\c{s}}}}, \bibinfo {author} {\bibfnamefont {E.}~\bibnamefont {Zhao}},\
  and\ \bibinfo {author} {\bibfnamefont {W.~V.}\ \bibnamefont {Liu}},\
  }\bibfield  {title} {\bibinfo {title} {Effective theory of interacting
  fermions in shaken square optical lattices},\ }\href@noop {} {\bibfield
  {journal} {\bibinfo  {journal} {Physical Review A}\ }\textbf {\bibinfo
  {volume} {95}},\ \bibinfo {pages} {063619} (\bibinfo {year}
  {2017})}\BibitemShut {NoStop}%
\bibitem [{\citenamefont {Zhang}\ \emph {et~al.}(2015)\citenamefont {Zhang},
  \citenamefont {Lang},\ and\ \citenamefont {Zhou}}]{zhang2015chiral}%
  \BibitemOpen
  \bibfield  {author} {\bibinfo {author} {\bibfnamefont {S.-L.}\ \bibnamefont
  {Zhang}}, \bibinfo {author} {\bibfnamefont {L.-J.}\ \bibnamefont {Lang}},\
  and\ \bibinfo {author} {\bibfnamefont {Q.}~\bibnamefont {Zhou}},\ }\bibfield
  {title} {\bibinfo {title} {Chiral d-wave superfluid in periodically driven
  lattices},\ }\href@noop {} {\bibfield  {journal} {\bibinfo  {journal}
  {Physical Review Letters}\ }\textbf {\bibinfo {volume} {115}},\ \bibinfo
  {pages} {225301} (\bibinfo {year} {2015})}\BibitemShut {NoStop}%
\bibitem [{\citenamefont {Kawamura}\ \emph {et~al.}(2022)\citenamefont
  {Kawamura}, \citenamefont {Ohashi},\ and\ \citenamefont
  {Hanai}}]{kawamura2022proposed}%
  \BibitemOpen
  \bibfield  {author} {\bibinfo {author} {\bibfnamefont {T.}~\bibnamefont
  {Kawamura}}, \bibinfo {author} {\bibfnamefont {Y.}~\bibnamefont {Ohashi}},\
  and\ \bibinfo {author} {\bibfnamefont {R.}~\bibnamefont {Hanai}},\ }\bibfield
   {title} {\bibinfo {title} {Proposed fermi-surface reservoir engineering and
  application to realizing unconventional fermi superfluids in a
  driven-dissipative nonequilibrium fermi gas},\ }\href@noop {} {\bibfield
  {journal} {\bibinfo  {journal} {Physical Review A}\ }\textbf {\bibinfo
  {volume} {106}},\ \bibinfo {pages} {013311} (\bibinfo {year}
  {2022})}\BibitemShut {NoStop}%
\bibitem [{\citenamefont {Zheng}\ \emph {et~al.}(2015)\citenamefont {Zheng},
  \citenamefont {Qu}, \citenamefont {Zou},\ and\ \citenamefont
  {Zhang}}]{zheng2015floquet}%
  \BibitemOpen
  \bibfield  {author} {\bibinfo {author} {\bibfnamefont {Z.}~\bibnamefont
  {Zheng}}, \bibinfo {author} {\bibfnamefont {C.}~\bibnamefont {Qu}}, \bibinfo
  {author} {\bibfnamefont {X.}~\bibnamefont {Zou}},\ and\ \bibinfo {author}
  {\bibfnamefont {C.}~\bibnamefont {Zhang}},\ }\bibfield  {title} {\bibinfo
  {title} {Floquet fulde-ferrell-larkin-ovchinnikov superfluids and majorana
  fermions in a shaken fermionic optical lattice},\ }\href@noop {} {\bibfield
  {journal} {\bibinfo  {journal} {Physical Review A}\ }\textbf {\bibinfo
  {volume} {91}},\ \bibinfo {pages} {063626} (\bibinfo {year}
  {2015})}\BibitemShut {NoStop}%
\bibitem [{\citenamefont {Zheng}\ \emph {et~al.}(2016)\citenamefont {Zheng},
  \citenamefont {Qu}, \citenamefont {Zou},\ and\ \citenamefont
  {Zhang}}]{zheng2016fulde}%
  \BibitemOpen
  \bibfield  {author} {\bibinfo {author} {\bibfnamefont {Z.}~\bibnamefont
  {Zheng}}, \bibinfo {author} {\bibfnamefont {C.}~\bibnamefont {Qu}}, \bibinfo
  {author} {\bibfnamefont {X.}~\bibnamefont {Zou}},\ and\ \bibinfo {author}
  {\bibfnamefont {C.}~\bibnamefont {Zhang}},\ }\bibfield  {title} {\bibinfo
  {title} {Fulde-ferrell superfluids without spin imbalance in driven optical
  lattices},\ }\href@noop {} {\bibfield  {journal} {\bibinfo  {journal}
  {Physical review letters}\ }\textbf {\bibinfo {volume} {116}},\ \bibinfo
  {pages} {120403} (\bibinfo {year} {2016})}\BibitemShut {NoStop}%
\bibitem [{\citenamefont {Kinnunen}\ \emph {et~al.}(2018)\citenamefont
  {Kinnunen}, \citenamefont {Baarsma}, \citenamefont {Martikainen},\ and\
  \citenamefont {T{\"o}rm{\"a}}}]{kinnunen2018fulde}%
  \BibitemOpen
  \bibfield  {author} {\bibinfo {author} {\bibfnamefont {J.~J.}\ \bibnamefont
  {Kinnunen}}, \bibinfo {author} {\bibfnamefont {J.~E.}\ \bibnamefont
  {Baarsma}}, \bibinfo {author} {\bibfnamefont {J.-P.}\ \bibnamefont
  {Martikainen}},\ and\ \bibinfo {author} {\bibfnamefont {P.}~\bibnamefont
  {T{\"o}rm{\"a}}},\ }\bibfield  {title} {\bibinfo {title} {The
  fulde--ferrell--larkin--ovchinnikov state for ultracold fermions in lattice
  and harmonic potentials: a review},\ }\href@noop {} {\bibfield  {journal}
  {\bibinfo  {journal} {Reports on Progress in Physics}\ }\textbf {\bibinfo
  {volume} {81}},\ \bibinfo {pages} {046401} (\bibinfo {year}
  {2018})}\BibitemShut {NoStop}%
\bibitem [{\citenamefont {Liao}\ \emph {et~al.}(2010)\citenamefont {Liao},
  \citenamefont {Rittner}, \citenamefont {Paprotta}, \citenamefont {Li},
  \citenamefont {Partridge}, \citenamefont {Hulet}, \citenamefont {Baur},\ and\
  \citenamefont {Mueller}}]{liao2010spin}%
  \BibitemOpen
  \bibfield  {author} {\bibinfo {author} {\bibfnamefont {Y.-a.}\ \bibnamefont
  {Liao}}, \bibinfo {author} {\bibfnamefont {A.~S.~C.}\ \bibnamefont
  {Rittner}}, \bibinfo {author} {\bibfnamefont {T.}~\bibnamefont {Paprotta}},
  \bibinfo {author} {\bibfnamefont {W.}~\bibnamefont {Li}}, \bibinfo {author}
  {\bibfnamefont {G.~B.}\ \bibnamefont {Partridge}}, \bibinfo {author}
  {\bibfnamefont {R.~G.}\ \bibnamefont {Hulet}}, \bibinfo {author}
  {\bibfnamefont {S.~K.}\ \bibnamefont {Baur}},\ and\ \bibinfo {author}
  {\bibfnamefont {E.~J.}\ \bibnamefont {Mueller}},\ }\bibfield  {title}
  {\bibinfo {title} {Spin-imbalance in a one-dimensional fermi gas},\ }\href
  {https://doi.org/10.1038/nature09393} {\bibfield  {journal} {\bibinfo
  {journal} {Nature}\ }\textbf {\bibinfo {volume} {467}},\ \bibinfo {pages}
  {567} (\bibinfo {year} {2010})}\BibitemShut {NoStop}%
\bibitem [{\citenamefont {Schneider}\ \emph {et~al.}(2008)\citenamefont
  {Schneider}, \citenamefont {Hackerm{\"u}ller}, \citenamefont {Will},
  \citenamefont {Best}, \citenamefont {Bloch}, \citenamefont {Costi},
  \citenamefont {Helmes}, \citenamefont {Rasch},\ and\ \citenamefont
  {Rosch}}]{partridge2006pairing}%
  \BibitemOpen
  \bibfield  {author} {\bibinfo {author} {\bibfnamefont {U.}~\bibnamefont
  {Schneider}}, \bibinfo {author} {\bibfnamefont {L.}~\bibnamefont
  {Hackerm{\"u}ller}}, \bibinfo {author} {\bibfnamefont {S.}~\bibnamefont {Will}},
  \bibinfo {author} {\bibfnamefont {T.}~\bibnamefont {Best}}, \bibinfo {author}
  {\bibfnamefont {I.}~\bibnamefont {Bloch}}, \bibinfo {author} {\bibfnamefont
  {T.~A.}\ \bibnamefont {Costi}}, \bibinfo {author} {\bibfnamefont {R.~W.}\
  \bibnamefont {Helmes}}, \bibinfo {author} {\bibfnamefont {D.}~\bibnamefont
  {Rasch}},\ and\ \bibinfo {author} {\bibfnamefont {A.}~\bibnamefont {Rosch}},\
  }\bibfield  {title} {\bibinfo {title} {Metallic and insulating phases of
  repulsively interacting fermions in a 3d optical lattice},\ }\href
  {https://doi.org/10.1126/science.1165449} {\bibfield  {journal} {\bibinfo
  {journal} {Science}\ }\textbf {\bibinfo {volume} {322}},\ \bibinfo {pages}
  {1520} (\bibinfo {year} {2008})},\ \Eprint
  {https://arxiv.org/abs/https://www.science.org/doi/pdf/10.1126/science.1165449}
  {https://www.science.org/doi/pdf/10.1126/science.1165449} \BibitemShut
  {NoStop}%
\bibitem [{\citenamefont {Olsen}\ \emph {et~al.}(2015)\citenamefont {Olsen},
  \citenamefont {Revelle}, \citenamefont {Fry}, \citenamefont {Sheehy},\ and\
  \citenamefont {Hulet}}]{olsen2015phase}%
  \BibitemOpen
  \bibfield  {author} {\bibinfo {author} {\bibfnamefont {B.~A.}\ \bibnamefont
  {Olsen}}, \bibinfo {author} {\bibfnamefont {M.~C.}\ \bibnamefont {Revelle}},
  \bibinfo {author} {\bibfnamefont {J.~A.}\ \bibnamefont {Fry}}, \bibinfo
  {author} {\bibfnamefont {D.~E.}\ \bibnamefont {Sheehy}},\ and\ \bibinfo
  {author} {\bibfnamefont {R.~G.}\ \bibnamefont {Hulet}},\ }\bibfield  {title}
  {\bibinfo {title} {Phase diagram of a strongly interacting spin-imbalanced
  fermi gas},\ }\href {https://doi.org/10.1103/PhysRevA.92.063616} {\bibfield
  {journal} {\bibinfo  {journal} {Phys. Rev. A}\ }\textbf {\bibinfo {volume}
  {92}},\ \bibinfo {pages} {063616} (\bibinfo {year} {2015})}\BibitemShut
  {NoStop}%
\bibitem [{\citenamefont {Long}\ \emph {et~al.}(2018)\citenamefont {Long},
  \citenamefont {Xiong}, \citenamefont {Gaire}, \citenamefont {Caligan},\ and\
  \citenamefont {Parker}}]{long2018all}%
  \BibitemOpen
  \bibfield  {author} {\bibinfo {author} {\bibfnamefont {Y.}~\bibnamefont
  {Long}}, \bibinfo {author} {\bibfnamefont {F.}~\bibnamefont {Xiong}},
  \bibinfo {author} {\bibfnamefont {V.}~\bibnamefont {Gaire}}, \bibinfo
  {author} {\bibfnamefont {C.}~\bibnamefont {Caligan}},\ and\ \bibinfo {author}
  {\bibfnamefont {C.~V.}\ \bibnamefont {Parker}},\ }\bibfield  {title}
  {\bibinfo {title} {All-optical production of li 6 molecular bose-einstein
  condensates in excited hyperfine levels},\ }\href@noop {} {\bibfield
  {journal} {\bibinfo  {journal} {Physical Review A}\ }\textbf {\bibinfo
  {volume} {98}},\ \bibinfo {pages} {043626} (\bibinfo {year}
  {2018})}\BibitemShut {NoStop}%
\bibitem [{\citenamefont {Long}\ \emph {et~al.}(2021)\citenamefont {Long},
  \citenamefont {Xiong},\ and\ \citenamefont {Parker}}]{long2021spin}%
  \BibitemOpen
  \bibfield  {author} {\bibinfo {author} {\bibfnamefont {Y.}~\bibnamefont
  {Long}}, \bibinfo {author} {\bibfnamefont {F.}~\bibnamefont {Xiong}},\ and\
  \bibinfo {author} {\bibfnamefont {C.~V.}\ \bibnamefont {Parker}},\ }\bibfield
   {title} {\bibinfo {title} {Spin susceptibility above the superfluid onset in
  ultracold fermi gases},\ }\href@noop {} {\bibfield  {journal} {\bibinfo
  {journal} {Physical Review Letters}\ }\textbf {\bibinfo {volume} {126}},\
  \bibinfo {pages} {153402} (\bibinfo {year} {2021})}\BibitemShut {NoStop}%
\bibitem [{\citenamefont {Ran{\c{c}}on}\ and\ \citenamefont
  {Levin}(2014)}]{ranccon2014equilibrating}%
  \BibitemOpen
  \bibfield  {author} {\bibinfo {author} {\bibfnamefont {A.}~\bibnamefont
  {Ran{\c{c}}on}}\ and\ \bibinfo {author} {\bibfnamefont {K.}~\bibnamefont
  {Levin}},\ }\bibfield  {title} {\bibinfo {title} {Equilibrating dynamics in
  quenched bose gases: Characterizing multiple time regimes},\ }\href@noop {}
  {\bibfield  {journal} {\bibinfo  {journal} {Physical Review A}\ }\textbf
  {\bibinfo {volume} {90}},\ \bibinfo {pages} {021602} (\bibinfo {year}
  {2014})}\BibitemShut {NoStop}%
\bibitem [{\citenamefont {Choudhury}\ and\ \citenamefont
  {Mueller}(2014)}]{choudhury2014stability}%
  \BibitemOpen
  \bibfield  {author} {\bibinfo {author} {\bibfnamefont {S.}~\bibnamefont
  {Choudhury}}\ and\ \bibinfo {author} {\bibfnamefont {E.~J.}\ \bibnamefont
  {Mueller}},\ }\bibfield  {title} {\bibinfo {title} {Stability of a floquet
  bose-einstein condensate in a one-dimensional optical lattice},\ }\href@noop
  {} {\bibfield  {journal} {\bibinfo  {journal} {Physical Review A}\ }\textbf
  {\bibinfo {volume} {90}},\ \bibinfo {pages} {013621} (\bibinfo {year}
  {2014})}\BibitemShut {NoStop}%
\bibitem [{\citenamefont {Choudhury}\ and\ \citenamefont
  {Mueller}(2015{\natexlab{a}})}]{choudhury2015stability}%
  \BibitemOpen
  \bibfield  {author} {\bibinfo {author} {\bibfnamefont {S.}~\bibnamefont
  {Choudhury}}\ and\ \bibinfo {author} {\bibfnamefont {E.~J.}\ \bibnamefont
  {Mueller}},\ }\bibfield  {title} {\bibinfo {title} {Stability of a
  bose-einstein condensate in a driven optical lattice: Crossover between weak
  and tight transverse confinement},\ }\href@noop {} {\bibfield  {journal}
  {\bibinfo  {journal} {Physical Review A}\ }\textbf {\bibinfo {volume} {92}},\
  \bibinfo {pages} {063639} (\bibinfo {year} {2015}{\natexlab{a}})}\BibitemShut
  {NoStop}%
\bibitem [{\citenamefont {Choudhury}\ and\ \citenamefont
  {Mueller}(2015{\natexlab{b}})}]{choudhury2015transverse}%
  \BibitemOpen
  \bibfield  {author} {\bibinfo {author} {\bibfnamefont {S.}~\bibnamefont
  {Choudhury}}\ and\ \bibinfo {author} {\bibfnamefont {E.~J.}\ \bibnamefont
  {Mueller}},\ }\bibfield  {title} {\bibinfo {title} {Transverse collisional
  instabilities of a bose-einstein condensate in a driven one-dimensional
  lattice},\ }\href@noop {} {\bibfield  {journal} {\bibinfo  {journal}
  {Physical Review A}\ }\textbf {\bibinfo {volume} {91}},\ \bibinfo {pages}
  {023624} (\bibinfo {year} {2015}{\natexlab{b}})}\BibitemShut {NoStop}%
\end{thebibliography}
\end{document}


\preprint{APS/123-QED}

\title{Instability and Momentum Bifurcation of a molecular BEC \\ in a Shaken Lattice with Exotic Dispersion}

\begin{CJK*}{UTF8}{gbsn}
\author{Kaiyue Wang (王凯越), Feng Xiong (熊风), \\Yun Long (龙云), Yun Ma (马芸), Colin V. Parker}

\affiliation{
School of Physics, Georgia Institute of Technology, Atlanta, Georgia 30332, USA
}%

\date{\today}

\maketitle
\end{CJK*} 

\section{Shaking Scheme}

\begin{figure}[b]
    \centering
    \includegraphics[width=\linewidth]{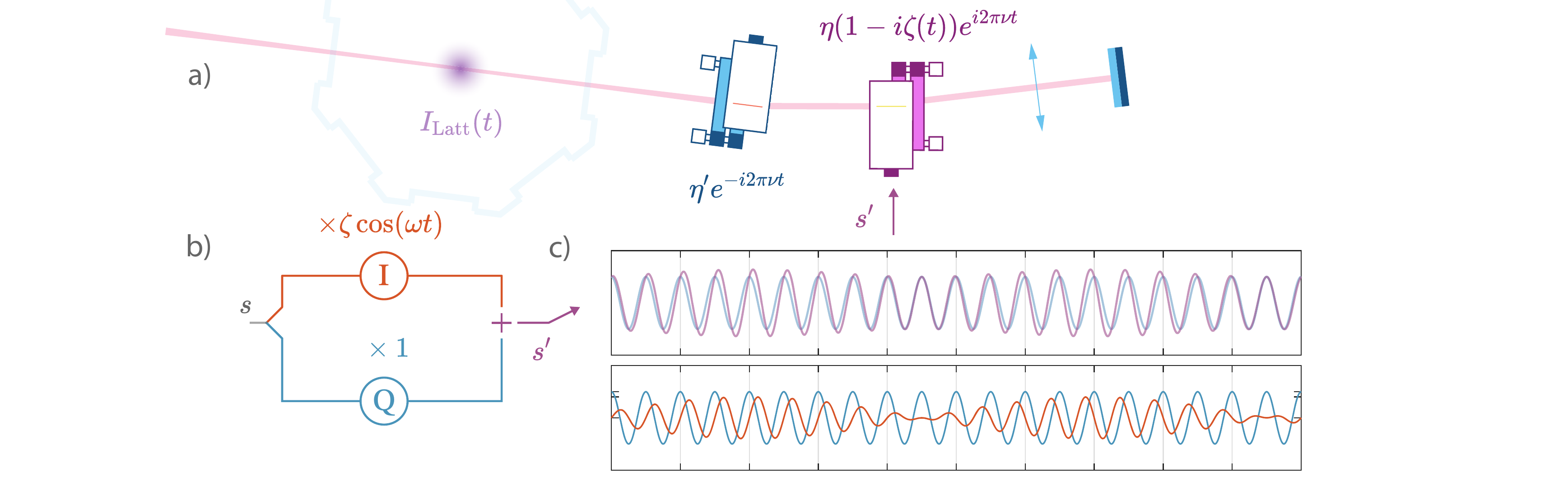}        
    \caption{Experimental configuration for lattice shaking with a doubled-passed AOM.}
    \label{fig:aom}
\end{figure}

One of the double-passed AOMs for the retro-reflected lattice beam has a modulated signal to create the shaking. The unmodified signal \(s = s_\text{max} \cos(2\pi\nu t)\) goes through an IQ modulator as shown in the diagram on the left of Fig. \ref{fig:aom}. The IQ modulator takes the signal and divides it into two components: an in-phase lattice component (created by a DC voltage on the``Q" channel of the IQ modulator), and an oscillating quadrature component that has the phase retarded by \(\pi/2\) (created by an AC voltage on the ``I" channel). Note that the usage of I and Q is reversed from typical configuration, although they are equivalent for our purposes. The input signal to the AOM, \(s'\), is the mixing of the two components with tunable amplitudes. We set the relative amplitude of Q to be \(1\), and I to be \(\zeta(t) = -\zeta_\text{max}\cos(\omega t)\). Therefore,
\begin{equation}
    s' = s_\text{max}\qty[1\cross\cos(2\pi\nu t) - \zeta(t) \cross \cos(2\pi\nu t - \dfrac{\pi}{2})],
\end{equation}
or in terms of complex amplitudes, 
\begin{equation}
    \tilde s' = \qty(1 - i\zeta(t)) \tilde s.
\end{equation}
In the right figure, the bottom diagram shows an exaggerated version of the two components (``I'' in red, ``Q'' in blue). The upper diagram shows the mixed signal \(s'\) in violet, and the unmodified \(s\) (blue) as a reference. As can be seen the modulation is in both phase and amplitude.

Consider the change of complex amplitude for the lattice light, which passes through the modulated AOM twice in the same order, and the unmodulated AOM (also with input frequency \(\nu\), in our case \(\nu = \SI{80}{MHz}\)) twice in the opposite order. For every pass of the modulated AOM, it gains a complex factor of \(\tilde\eta_+ = \eta \qty(1-i\zeta(t)) e^{-i 2\pi\nu t}\), and for the unmodulated AOM a factor of \(\tilde\eta_- = \eta' e^{i 2\pi\nu t}\). Here, \(\eta\) and \(\eta'\) are the real efficiencies for amplitudes. The retro-reflected beam in the chamber thus carries a factor of \(\tilde\eta = \tilde\eta_+^2\tilde\eta_-^2 = \eta^2\eta'^2 \qty(1 - 2i\zeta(t) - \zeta(t)^2)\). 

We now calculate the light intensity at the center of the chamber. Assuming the incoming light has a complex amplitude of \(\tilde A_\text{inc}(t) = A e^{ikz}e^{-ikct}\), the retro-reflected light has complex amplitude of \(\tilde A_\text{retro}(t) = A e^{-ikz}e^{-ikct}\tilde\eta\). The intensity is,
\begin{eqnarray}
    I(t) = \qty|\tilde A_\text{inc} + \tilde A_\text{retro}|^2 &=& \qty(\tilde A_\text{inc}^\ast + \tilde A_\text{retro}^\ast)\qty(\tilde A_\text{inc} + \tilde A_\text{retro}) \nonumber\\
    &=& \qty|\tilde A_\text{inc}|^2 + \qty|\tilde A_\text{retro}|^2 + 2\Re \tilde A_\text{inc} \tilde A_\text{retro}^\ast
\end{eqnarray}
Only the third term contains spatial modulations and serves as an optical lattice. The intensity is
\begin{eqnarray}
    I_\text{Latt}(t) &=& 2\Re \tilde A_\text{inc} \tilde A_\text{retro}^\ast = 2 A^2 \eta^2\eta'^2  \Re \qty[e^{2ikz} \qty(1-\zeta(t)^2 + i 2\zeta(t))] \nonumber \\
    &=& 2I_\text{max} \qty[\qty(1 - \zeta(t)^2) \cos(2kz) + 2\zeta(t)\cos(2kz + \frac{\pi}{2})] \nonumber \\
    &\approx& 2 I_\text{max} \qty[\cos(2kz) + \xi_\text{max}(t)\cos(2kz + \frac{\pi}{2})\cos(\omega t)],
\end{eqnarray}
corresponding to an in-phase lattice component with amplitude \(1 - \zeta(t)^2\) close to the unmodulated static lattice and a \(\pi/2\) shifted lattice component with amplitude \(2\zeta(t)\). At lower maximum signal amplitudes \(\zeta_\text{max}\), the shifted lattice oscillates at double the lattice amplitude: \(\xi_\text{max} = 2 \zeta_\text{max}\). The small modulation of lattice intensity at twice the shaking frequency can be neglected to a good approximation. For example, if \(\xi_\text{max} = 1/2\), then \(\zeta_\text{max} = 1/4\), and \(\zeta^2(t) = [1+\cos(2\omega t)]/32\).

\section{Sequence details}

Figure \ref{fig:seq} below shows the sequence of each experimental run. First, the incoming lattice beam is turned on adiabatically in \SI{99}{ms}. The mBEC's OD profile is shown on the top left of the diagram, and serves as the initial condition of the evolution, yielding the initial size parameters and \(\theta_V\). Second, the retro-reflected beam is turned on over \SI{1}{ms} by ramping up the signal to the pair of AOMs, and the incoming lattice beam is weakened to maintain the overall average intensity. Then, the IQ modulation that produces the shaking is ramped up over \SI{0.1}{\milli\second}, and the frequency of the IQ modulation is ramped down from \SI{80}{kHz} to the target \(f\) over another \SI{0.1}{\milli\second} (reflected as the tone change in the shading in the figure). This enables the mBEC to be loaded onto the desired hybridized band. Next, with constant shaking frequency and amplitude \(\omega\) and \(\xi_\text{max}\), the mBEC evolves for a controlled period of time \(t_\text{shake}\). Starting from the final \SI{5}{\milli\second} of the lattice beam turn-on, the magnetic field is kept at \SI{650}{G}, far on the BEC-side of resonance, where Feshbach molecules are bound. Finally, the destructive imaging pulse can be taken at multiple possible times. In-situ images are taken directly at the end of the shaking time. In order to minimize the motion of atoms during the pulse this requires the shortest possible pulse, which for our experiment is \SI{1}{\micro\second}, and to reduce shot noise the image is averaged over 100 shots. The imaging frequency is chosen by optimizing the optical density of the weakly-bound molecules, which may break apart during the pulse. For short TOF images, within \SI{0.1}{\milli\second} to \SI{1}{\milli\second} after release of the trap, a longer \SI{100}{\micro\second} imaging pulse can be used. For longer TOF images, we can afford to ramp the magnetic field closer to the resonance, where the Feshbach pairs are less tightly bound and easier to image. A long TOF image that resolves the momentum distribution is typically taken at \SI{3}{\milli\second}.

\begin{figure}[hb!]
    \centering
    \includegraphics[width=\linewidth]{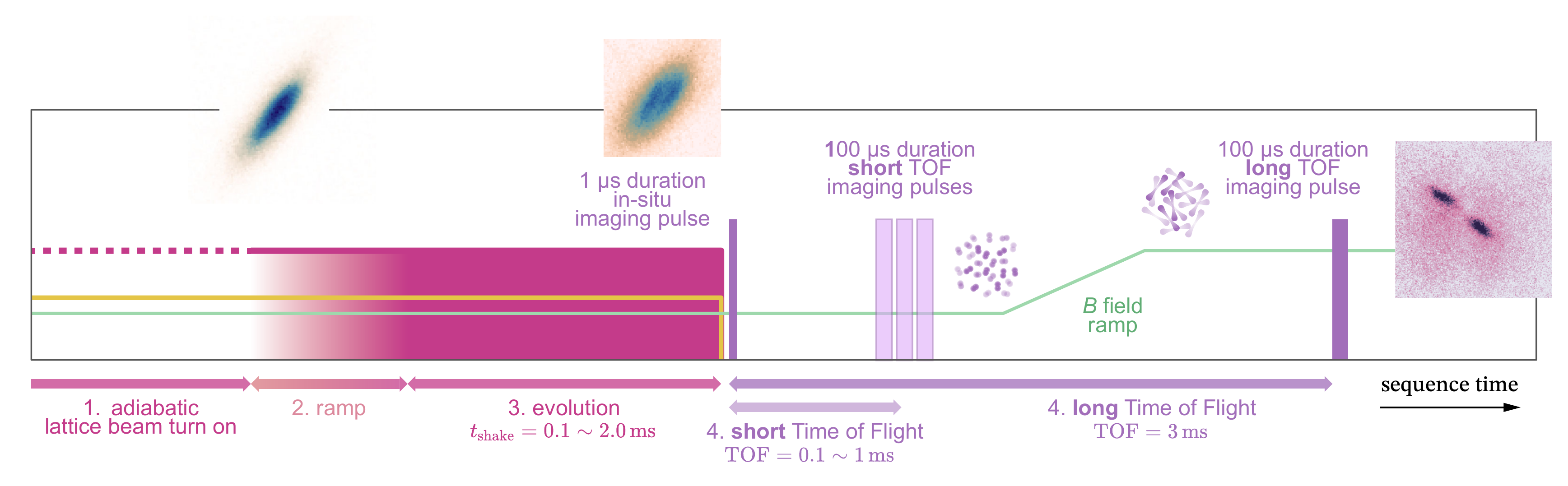}   
    \caption{Experimental sequence (see description in the supplemental text).}
    \label{fig:seq}
\end{figure}

\section{Extrapolation}

To reconstruct the in-situ distribution from short TOF images, we fit each TOF image with the sum of two 2D-Gaussian peaks, each of which is allowed to have an arbitrary rotation angle and asymmetric long/short radius:
\begin{eqnarray}
    f(R) &=& f_+(R) + f_-(R) \nonumber \\
    f_\pm(R) &=& A_\pm \exp{-\dfrac{1}{2} \qty(R - \mu_\pm)^T \Sigma_\pm^{-1} \qty(R - \mu_\pm)}
\end{eqnarray}
For conciseness we shall not label \(\pm\) when introducing the following symbols and focus on only one of the peaks. Here, \(R = \smallmat{x \\ z}\) is the 2D positional vector, \(\mu = \smallmat{\mu_x \\ \mu_z}\) is the center of the corresponding Gaussian peak, \(\Sigma^{-1} = U^T \smallmat{ \frac{1}{\sigma_1^2} & \\ & \frac{1}{\sigma_2^2} } U\) is the scaling matrix, where 
\(U = \smallmat{\cos\theta & -\sin\theta \\ \sin\theta & \cos\theta}\) 
is the rotation matrix, and \(\sigma_{1,2}\) are the two size parameters in length units, representing the Gaussian radius along the long and short axis. Of course, we need to make sure that the two peaks in each image are selected consistently. 

To extrapolate to the {\it in-situ} time \(t_0\) from other two known time points \(t_\alpha\) and \(t_\beta\), we assume the peaks' distribution follows an optimal transportation \cite{chen2018optimal}:
\begin{eqnarray}
    \Sigma(\tau) = \qty[(1 - \tau) \mathds{1} + \tau A_{\alpha;\beta}] \Sigma_\alpha \qty[(1 - \tau) \mathds{1} + \tau A_{\alpha;\beta}] 
\end{eqnarray}
Here, \(\tau = \dfrac{t - t_\alpha}{t_\beta - t_\alpha}\) is the normalized time, so that \(\Sigma(0) = \Sigma_\alpha\), \(\Sigma(1) = \Sigma_\beta\). and \(A_{\alpha;\beta}\) is the transportation matrix given by
\begin{eqnarray}
    A_{\alpha;\beta} = \Sigma_\alpha^{-\frac{1}{2}} \qty(\Sigma_\alpha^{\frac{1}{2}} \Sigma_\beta \Sigma_\alpha^{\frac{1}{2}} )^\frac{1}{2} \Sigma_\alpha^{-\frac{1}{2}} 
\end{eqnarray}
The inverse and square root of the \(\Sigma\) matrices are well-defined since they can be diagonalized through 2D-rotations. The peaks in the constructed in-situ image is acquired by choosing \(t_\alpha = \SI{0.8}{\milli\second}\) and \(t_\beta = \SI{2.5}{\milli\second}\), then taking in \(\tau = \dfrac{t_0 - t_\alpha}{t_\beta - t_\alpha}\).

\section{Simulation Scheme}

The simulation is run through a MATLAB application. The wavefunction is represented by a complex 3D-array \(\psi(x,y,z)\), typically of size \(100 \times 10 \times 100\). We choose the resolution in the lattice direction \(z\) to be exactly \(\lambda_\text{L}/2\), so that the discrete Fourier transform of the wavefunction fills one Brillouin zone along \(z\). At each step of evolution, we calculate the phase accumulation at each site by multiplying \((i - \Gamma)\qty(V_\text{trap}(x,y,z) + g\qty|\psi(x,y,z)|^2)\Delta t/\hbar\), then we transform \(\psi\) into Fourier space as \(\phi (k_x,k_y,q_z)\) and rotate in complex phase per wavevector site with the calculated dispersion \((i - \Gamma) D_{V_\text{L},\omega,\xi_\text{max}}(\bm{q} = (k_x,k_y,q_z)) \Delta t/\hbar\). The time step \(\Delta t\) is chosen to be \(0.02 \cdot h /E_\text{R}\) so that it is sufficiently small. The dissipation coefficient \(\Gamma\) is chosen to be 0.05 to best fit our experimental results. We record the \(\psi\) and \(\phi\) at selected time points for further analysis.

For each recorded frame, if the \(\phi\) indeed shows the two-cluster distribution, we acquire the trajectory parameters \(\Delta q\) and \(\theta_q\) by separating \(\phi = \phi_+ + \phi_-\), where \(\phi_\pm(q_z \lessgtr 0) = 0\). Then the center of mass of each segment is calculated. The limitation of this method is that in the case of small or no separation, a small non-zero value is still computed, but this does not prevent useful comparison of the simulation with the experiment.

\section{Initial Bifurcation Angle}
\begin{figure}[h!]
    \centering
    \includegraphics[width=\linewidth]{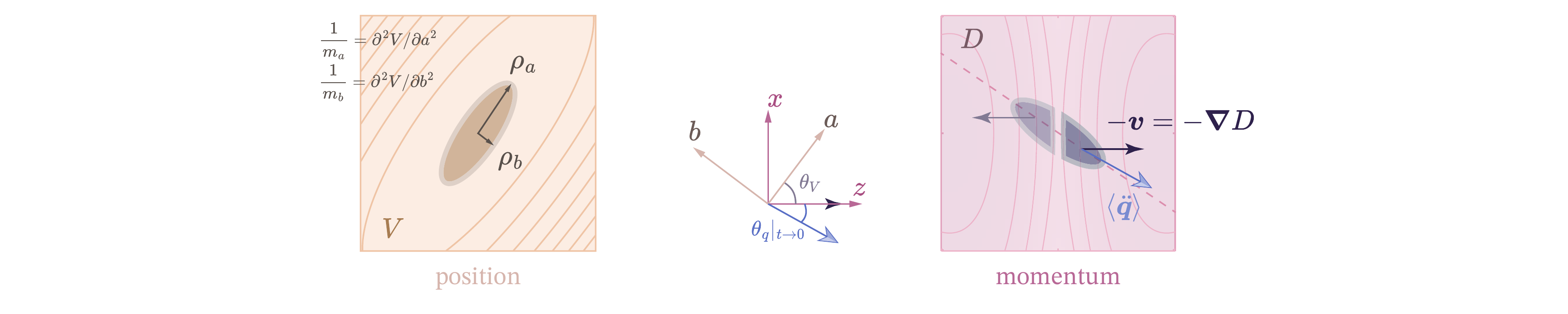}
    \caption{}
    \label{fig:geo}
\end{figure}
To understand the dynamics in the momentum space , it is more intuitive to reverse the role of momentum and position, i.e. to consider the trap potential \(V\) as a dispersion relation and the effective dispersion relation \(D\) as potentials. This is especially useful in our case since our effective dispersion is exotic while the trap potential is approximately quadratic, albeit anisotropic. For convenience, we introduce another set of base directions \(a, b\) that aligns with the trap potential's weakest and strongest confinement in the \(x, z\) plane, and the angle \(\theta_V\) is the angle between \(\hat{\bm{a}}\) and \(\hat{\bm{z}}\). The the effective force on the trajectory in momentum space, which is also the negative velocity in real space \(-\bm{v}_D = -\bm{\nabla}D\) is in the \(\pm z\) direction, for the two parts of the cluster lying on either side of the zone center respectively. Our goal is to calculate the direction of the initial acceleration in momentum space, \(\langle
\ddot{\bm q}\rangle\), under such force \(-\bm{v}_D\), where \(\bm q = (k_x, q_z)\) is the 2D momentum for the cluster, \(k_x\) and \(q_z\) are the wavevector in \(x\) and quasiwavevector in \(z\). See Fig. \ref{fig:geo}.

Taking the part of the cluster lying on the \(+z\) side as an example, we treat it as a classical particle with Hamiltonian in \(a, b\) basis
\begin{equation}
    \mathcal{H}(a, b, \bm q) = \dfrac{a^2}{2m_a} + \dfrac{b^2}{2m_b} + D(\bm q) ,
\end{equation}
The ratio between the effective masses \(m_a=1/\pdv[2]{V}{a}\) and \(m_b=1/\pdv[2]{V}{b}\) is determined by the elongation of the trap \(V\): \(m_a/m_b = (\rho_a/\rho_b)^2 = \kappa_V\) (We use \(\rho\) to mark on the diagram the distance from the trap center to the equipotential lines in the \(a,b\) direction, but the ratio is actually yielded through trap frequencies). 
This gives an equation of motion
\begin{eqnarray}
    \ddot{\bm q} &=& -\bm{\nabla}D \cdot \qty(\dfrac{\hat{\bm a}}{m_a} + \dfrac{\hat{\bm b}}{m_b}) \nonumber \\
    &=& -\dfrac{v_D}{m_a} \qty(-\kappa_V\sin\theta_V \hat{\bm b} + \cos\theta_V \hat{\bm a}) \nonumber \\
    &=& -\dfrac{v_D}{m_a} \qty[-(\kappa_V-1)\sin\theta_V\cos\theta_V \hat{\bm x} + (\cos^2\theta_V+\kappa_V\sin^2\theta_V) \hat{\bm z}] \nonumber
\end{eqnarray}
Therefore, the initial angle the bifurcation is \(\theta_q|_{t\rightarrow 0} = \arctan\abs{\dfrac{(\kappa_V-1)\sin\theta_V\cos\theta_V}{\cos^2\theta_V+\kappa_V\sin^2\theta_V}}\)
\section{Alternative Cases}
\begin{figure}[h!]
    \centering
    \includegraphics[width=\linewidth]{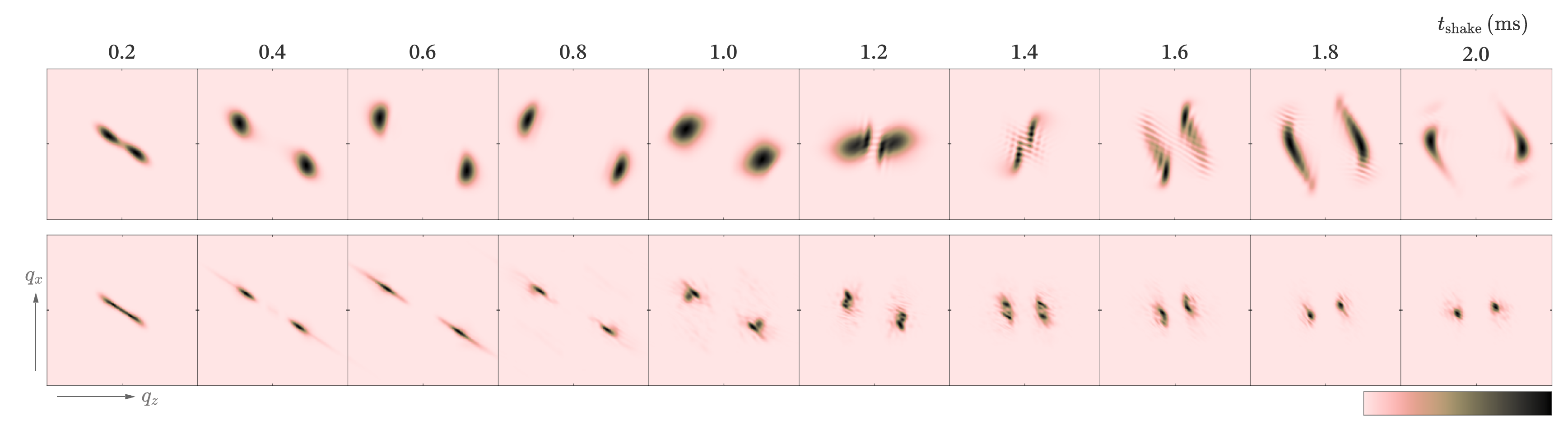}
    \caption{Comparison between the interacting and non-interacting trajectory from simulation.}
    \label{fig:ntrc}
\end{figure}
We demonstrate in Fig. \ref{fig:ntrc} the non-interacting case from the simulation (top) and compare it to one with interaction strength that matches our experiment (bottom). It's easy to see that the exotic dispersion makes the non-interactive case much more dispersive while in the interacting case the clusters are much more concentrated.
\begin{figure}[h!]
    \centering

    \includegraphics[width=\linewidth]{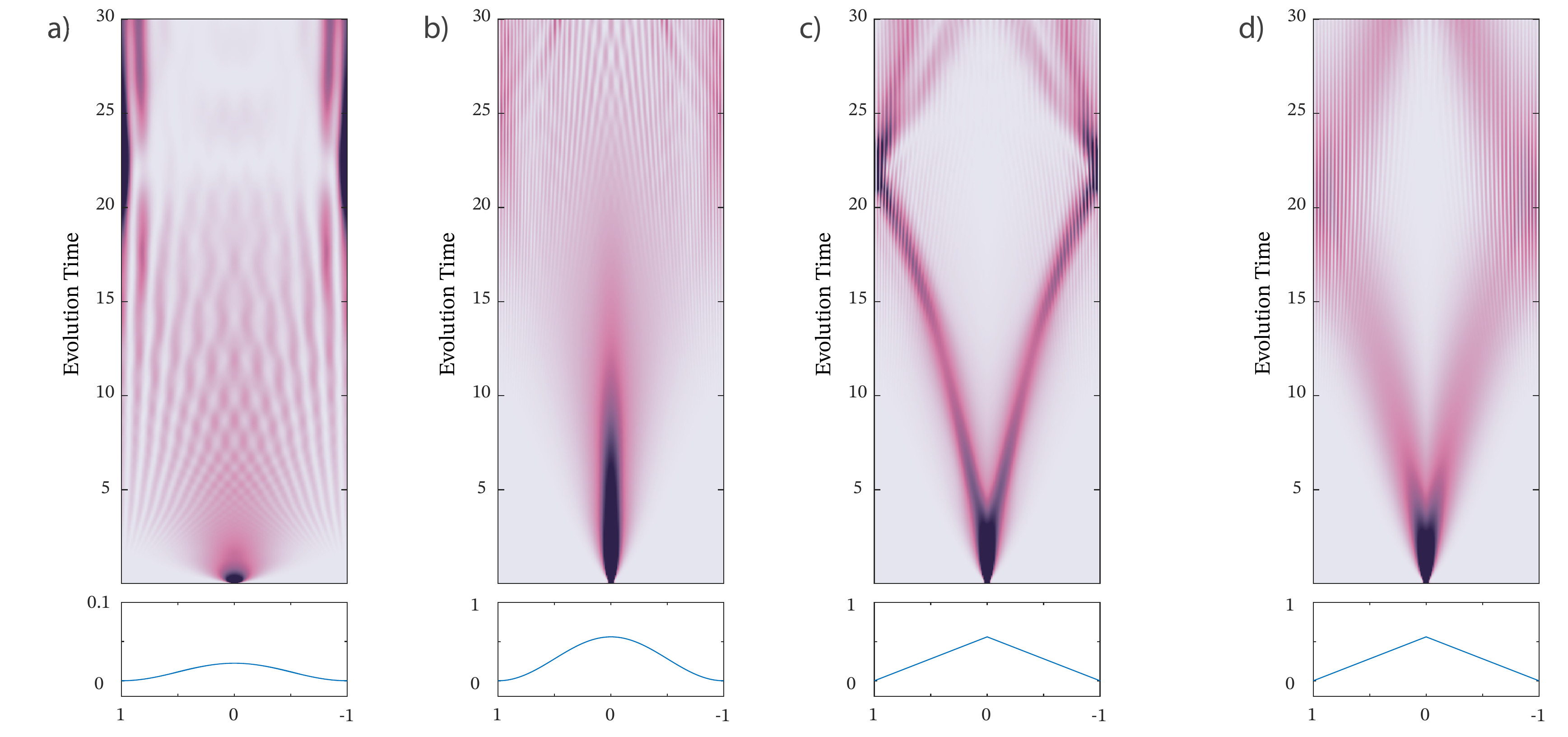}
    \caption{Comparisons between four simulation cases with dispersion minima at the zone edges. In each sub figure: the main diagram shows the evolution of the distribution over the quasimomentum space in \(z\) direction. The bottom panel shows the dispersion relation along \(z\) axis. \textbf{b,c,d} uses a larger scale for the dispersion curve.}
    \label{fig:cmpr}
\end{figure}
To compare with the results from \cite{mitchell2021floquet}, we simulate using a dispersion that has a single minimum at the zone edges, and \(\theta_V = 0\). Fig. \ref{fig:cmpr}a shows the case where the initial dispersion is much weaker than the potential and features the incoherent decay pattern. In Fig. \ref{fig:cmpr}b,c the dispersion is multiplied by 10 while the potential energy is weakened to \(1/9\), but in Fig. \ref{fig:cmpr}c the dispersion has a sharp peak at the maxima, which yields decay to the zone center in the form of solitons. Fig. \ref{fig:cmpr}d shows a similar case to Fig. \ref{fig:cmpr}c with no interaction, with a dispersive decay.

%